\documentclass[12pt,a4paper,twoside]{article}

\def\TheAuthor{Simon Trebst}                                   
\def\TheTitle{Kitaev Materials}                        
\def\TheHeading{Kitaev Materials}                                
\def\TheInstitute{Institute for Theoretical Physics}              
\def\TheAddress{University of Cologne, 50937 Cologne, Germany}            
\def\ThePart{}                                             
\def\TheChapter{}                                          

\usepackage{ferienschule_15}                                  

\makeindex

\newcommand{\ie}{{i.e.~}}
\newcommand{\eg}{{e.g.~}}

\newcommand{\Z}{Z$_{\rm 2}$}

\newcommand{\NaIrO}{Na$_{\text 2}$IrO$_{\text 3}$}
\newcommand{\LiIrO}{Li$_{\text 2}$IrO$_{\text 3}$}
\newcommand{\aLiIrO}{$\alpha$-Li$_{\text 2}$IrO$_{\text 3}$}
\newcommand{\bLiIrO}{$\beta$-Li$_{\text 2}$IrO$_{\text 3}$}
\newcommand{\cLiIrO}{$\gamma$-Li$_{\text 2}$IrO$_{\text 3}$}
\newcommand{\RuCl}{$\alpha$-RuCl$_{\text 3}$}
\newcommand{\bRuCl}{$\beta$-RuCl$_{\text 3}$}

\newcommand{\SrIrO}{Sr$_{\text 2}$IrO$_{\text 4}$}

\begin{document}

\MakeTitel           
\tableofcontents     

\vfill
\rule{42mm}{0.5pt}\\
{\footnotesize Lecture Notes of the $48^{{\rm th}}$ IFF Spring
School ``Topological Matter -- Topological Insulators, Skyrmions \newline and Majoranas''
(Forschungszentrum J{\"{u}}lich, 2017). All rights reserved. }

\newpage


\section{Spin-orbit entangled Mott insulators}

\begin{figure}[h]
    \centering
     \includegraphics[width=0.8\hsize]{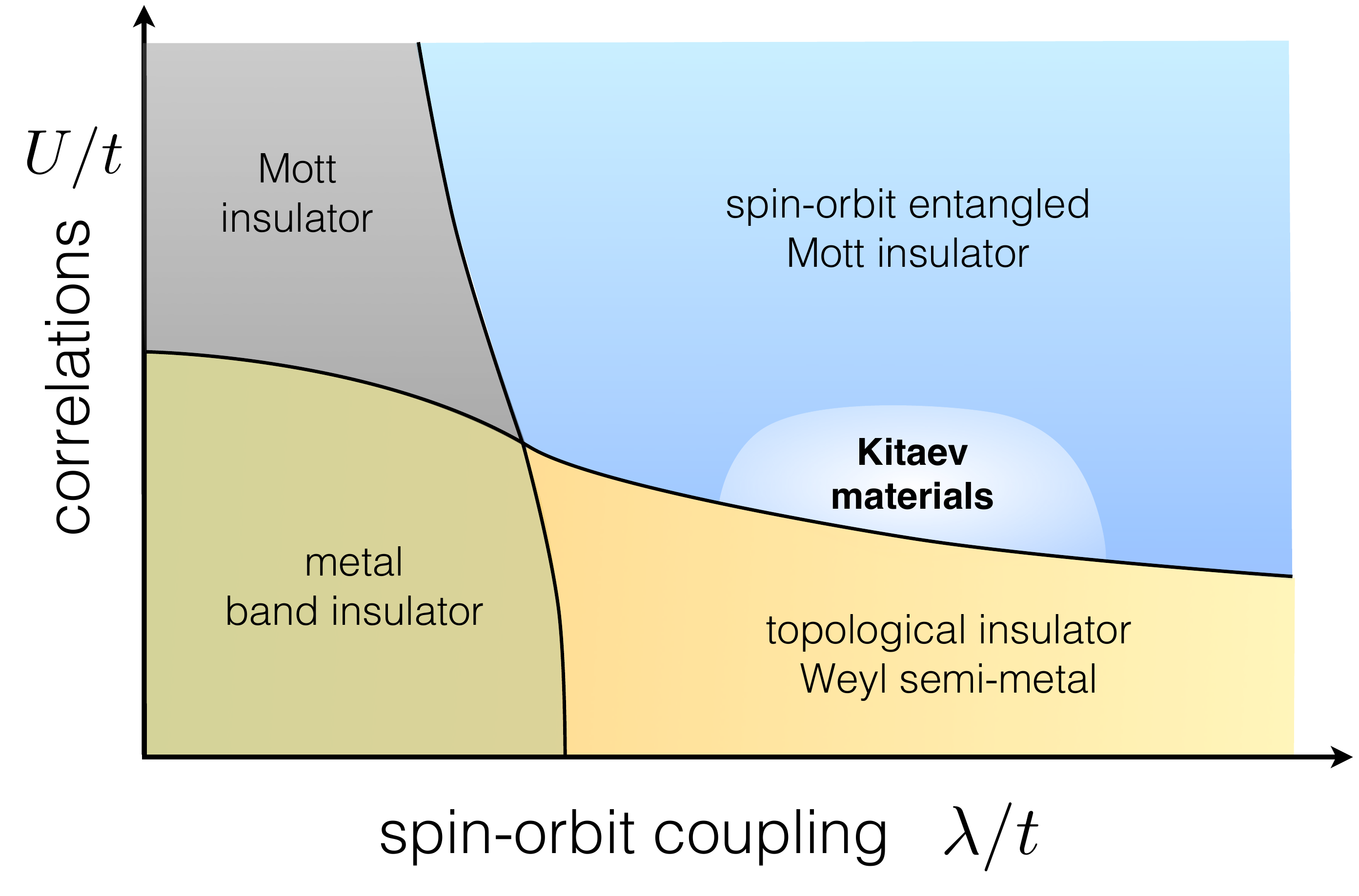}
     \caption{General form of the phase diagram in the presence of electronic correlations and spin-orbit coupling. 
     		   Figure adapted from the review~\cite{WitczakKrempa2014}.}
     \label{Fig:GeneralPhaseDiagram}
\end{figure}

Transition-metal oxides with partially filled $4d$ and $5d$ shells exhibit an intricate interplay of electronic, spin, and orbital degrees of freedom arising from a largely accidental balance of electronic correlations, spin-orbit entanglement, and crystal-field effects \cite{Pesin2010}. With different materials exhibiting slight tilts towards one of the three effects, a remarkably broad variety of novel forms of quantum matter can be explored. On the theoretical side, topology is found to play a crucial role in these systems -- an observation which, in the  weakly correlated regime, has lead to the discovery of the topological band insulator \cite{Hasan2010,Qi2011} and subsequently its metallic cousin, the Weyl semi-metal \cite{Wan2011,Yan2017}. Upon increasing electronic correlations, Mott insulators with unusual local moments such as spin-orbit entangled degrees of freedom can form and whose collective behavior gives rise to unconventional types of magnetism including the formation of quadrupolar correlations or the emergence of so-called spin liquid states. A rough guide to these novel types of quantum matter, currently widely explored in materials with substantial spin-orbit coupling, is given by the general phase diagram of Fig.~\ref{Fig:GeneralPhaseDiagram}, adapted from an early review~\cite{WitczakKrempa2014} of this rapidly evolving field at the current forefront of condensed matter physics.

\begin{figure}[t]
    \centering
     \includegraphics[width=\hsize]{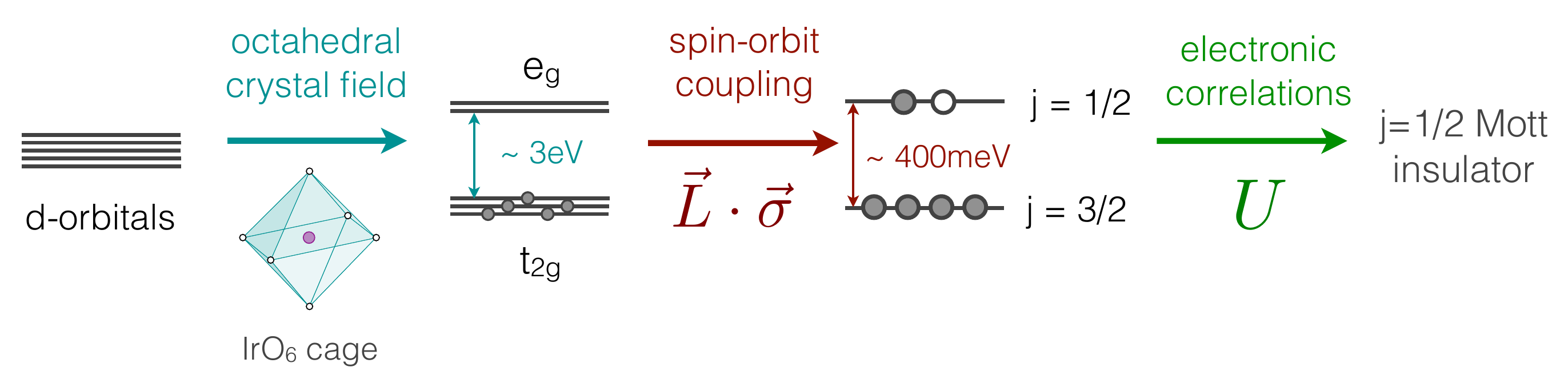}
     \caption{Formation of spin-orbit entangled $j=1/2$ moments for  ions in a $d^{\text 5}$ electronic configuration 
     		  such as for the typical iridium valence Ir$^{\text 4+}$ or the ruthenium valence Ru$^{\text 3+}$.}
     \label{Fig:SpinOrbitEntangledMomenta}
\end{figure}

Our focus here will be on the particularly intriguing scenario of the formation of novel types of Mott insulators in which the local moments are spin-orbit entangled $j=1/2$ Kramers doublets \cite{Khaliullin2005,Chen2008,Jackeli2009}. The latter are formed for ions in a $d^{\text 5}$ electronic configuration -- a vast family of such $d^5$ materials exists that not only includes most iridates, which typically have a Ir$^{4+}$ $(5d^5)$ valence, but also some Ru-based materials with a Ru$^{3+}$ $(4d^5)$ valence along with the (extremely toxic) osmates and (so far largely unexplored) rhenates.
The level scheme of such $d^{\text 5}$ ionic systems is illustrated in Fig.~\ref{Fig:SpinOrbitEntangledMomenta}. With the crystal field (e.g.\ of an octahedral oxygen cage) splitting off the two $e_{\text g}$ levels, this puts the five electrons with a total $s = 1/2$ magnetic moment into the $t_{\text 2g}$ orbitals with an effective $\ell = 1$ orbital moment. Strong spin-orbit coupling then results in a system with a fully filled $j = 3/2$ band and a half-filled $j = 1/2$ band. The reduced bandwidth of the latter then allows for the opening of a Mott gap even for the relatively moderate electronic correlations of the $4d$ and $5d$ compounds. This latter point should again be emphasized. The formation of a Mott insulator in these compounds is per se somewhat counterintuitive as the larger atomic radii of their $4d$ and $5d$ constituents give rise to considerable atomic overlap, which results in a large electronic bandwidth and thus an effective suppression of the electronic correlations. As such, conventional wisdom has long considered these ``heavy'' transition metal compounds to generically form metallic states. It is only because of an enormously enhanced spin-orbit coupling that the effective electronic bandwidth of these materials can be reduced (via the formation of two separate $j=3/2$ and $j=1/2$ bands) to a level that the largely suppressed electronic correlations can still drive the system into a Mott insulating state. These $j=1/2$ Mott materials are therefore sometimes referred to as ``spin-orbit assisted Mott insulators" and are located in the proximity of the metal-insulator transition for large spin-orbit coupling in the general phase diagram of Fig.~\ref{Fig:GeneralPhaseDiagram}.

The formation of such a $j=1/2$ Mott insulator was  first observed experimentally in 2008/2009
\cite{Kim2008,Kim2009} for the perovskite iridate \SrIrO\ --  a $5d$ transition-metal oxide which is  an isostructural analogue of La$_2$CuO$_4$, the parent compound of the cuprate superconductors. Remarkably, its low-temperature physics indeed shares a striking resemblance to the phenomenology of the cuprate superconductors including the formation of long-range antiferromagnetic  order of the $j=1/2$ pseudospins for the undoped material and the emergence of a pseudogap phase and associated Fermi arcs for electron-doped systems \cite{Kim2014,Torre2015,Cao2016a}. There is an intense ongoing search \cite{Kim2016d,Yan2015} for superconductivity in \SrIrO\ and other perovskite iridates.

In parallel, much attention has been drawn towards $j=1/2$ Mott insulators that exhibit bond-directional exchange interactions and are thought to exhibit unconventional forms of magnetism, such as the emergence of spin liquids \cite{Balents2010,Savary2017} or the formation of non-trivial spin textures.  We refer to these systems as {\em Kitaev materials}.
Of particular interest here are the sister compounds \NaIrO\ and \aLiIrO\ and more recently \RuCl\ that form Mott insulators, in which local $j=1/2$ moments are aligned in (almost decoupled) hexagonal layers. As such they are perfect candidate materials for a solid-state
realization of the Kitaev honeycomb model \cite{Kitaev2006} as envisioned by Jackeli and Khaliullin in 2009 \cite{Jackeli2009}. 
Over the last few years, materials synthesis of novel $j=1/2$ Mott insulators has been thriving at a remarkable pace and has further broadened our view on Kitaev materials beyond the original honeycomb structure. As will be outlined in the remainder of this chapter, 
$j=1/2$ Mott insulators with strong bond-directional exchange interactions have also been synthesized for triangular lattice geometries, e.g. in the family of hexagonal perovskites Ba$_3$Ir$_x$Ti$_{3-x}$O$_9$, and three-dimensional, tricoordinated lattice generalizations of the honeycomb lattice (dubbed the hyper-honeycomb and stripy-honeycomb) in polymorphs of \aLiIrO, \ie the iridates \bLiIrO\ and \cLiIrO, respectively. In addition, theoretical progress has been made in comprehensively classifying the gapless spin liquid ground states of three-dimensional Kitaev models. It will be the purpose of this chapter to provide what is hopefully a concise review of the current status of these Kitaev materials and their conceptual understanding.
%


\subsection{Bond-directional interactions}

At the heart of all Kitaev materials are bond-directional interactions, \ie Ising-like interactions where the exchange 
easy axis depends on the spatial orientation of an exchange bond, and which dominate in coupling strength over all other exchange types. 
The microscopic origin of such bond-directional interactions in $d^5$ transition metal compounds has been worked out in a pioneering 2005 paper by Khaliullin \cite{Khaliullin2005} and later refined to the context of Kitaev-type interactions in joint work of Jackeli and Khaliullin \cite{Jackeli2009}. These papers have undoubtedly shaped the formation of what is now a burgeoning field of experimental and theoretical exploration of Kitaev materials. 

\begin{figure}[h]
    \centering
     \includegraphics[width=\hsize]{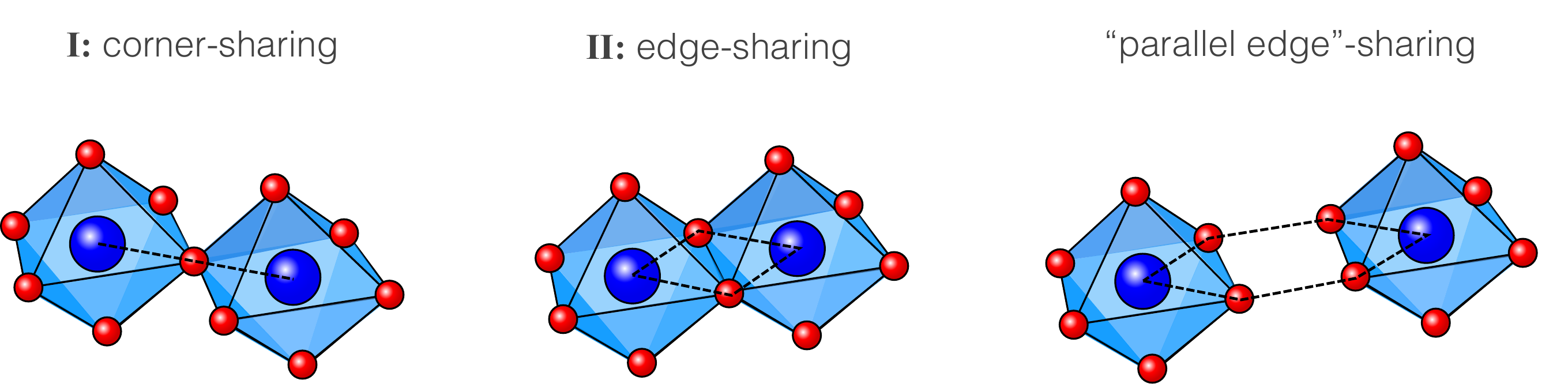}
     \caption{Illustration of possible geometric orientations of neighboring IrO$_6$ octahedra that give rise to different types 
     		of (dominant) exchange interactions between the magnetic moments located on the iridium ion at the center of these octahedra.
		For the corner-sharing geometry (I) one finds a dominant symmetric Heisenberg exchange, 
		while for the edge-sharing geometries (II) one finds a dominant bond-directional, Kitaev-type exchange.
         	}
     \label{Fig:IridateExchangePaths}
\end{figure}

What Khaliullin and Jackeli have realized is that the geometric orientation of neighboring IrO$_6$ octahedra plays a crucial role in
determining the microscopic exchange of the magnetic moments located on the iridium ion at the center of these octahedra. They distinguish two elementary scenarios, which are illustrated (along with an extension) in Fig.~\ref{Fig:IridateExchangePaths}. First, for perovskite iridates such as \SrIrO\ two neighboring IrO$_6$ octahedra share a {\em corner}. For the coupling between the two neighboring iridium ions this means that there is a single Ir-O-Ir exchange path, which is also referred to as 180$^\circ$ bond. The dominant coupling along this type of bond is -- despite the presence of a strong spin-orbit coupling -- a symmetric Heisenberg exchange between the spin-orbit entangled $j=1/2$ moments. 
The second scenario plays out in materials, in which two neighboring IrO$_6$ octahedra share an {\em edge}. Here there are two Ir-O-Ir exchange paths, which exhibit a  90$^\circ$ bonding geometry. The fact that there are {\em two} exchange paths turns out to be crucial, as the two alternative paths lead to a destructive interference of the symmetric Heisenberg exchange when restricting the coupling to arise exclusively from the $j=1/2$ bands, or alternatively to a significant suppression to some residual Heisenberg exchange when considering the full multi-orbital model also including the (virtual) $j=3/2$ bands. In lieu of the highly suppressed isotropic exchange, it is a bond-directional coupling, stemming from Hund's coupling and mediated through the multiplet structure of the excited levels, that takes center stage and becomes the dominant coupling. The bond-directionality of the coupling arises since the pair of $d$-orbitals linked for two neighboring octahedra depends on the type of edge shared at their intersection. This in turn gives rise to a spatially oriented Ising-type coupling between the spin-orbit entangled $j=1/2$ moments where the magnetic easy-axis is perpendicular to the plane spanned by the two exchange paths \cite{Khaliullin2005,Jackeli2009}. The strength of this bond-directional coupling is given by
\begin{equation}
	~ - \frac{8 t^2 J_H}{3 U^2} S_1^{\gamma} S_2^{\gamma} \,,
\end{equation}
where $t$ is the effective interorbital hopping mediated by the oxygen ions, $J_H$ is the Hund's coupling and $U$ the electronic correlation strength. The two spin-orbit entangled $j=1/2$ moments, represented by the usual vectors of SU(2) spin-1/2 Pauli matrices ${\bf S}_1$ and ${\bf S}_2$, are coupled only via a single spin component $\gamma = x,y,z$.
As such the edge-sharing 90$^\circ$ bond geometry naturally gives rise to a quantum {\em compass model} \cite{Nussinov2015}, originally introduced by Kugel and Khomskii for the orbital degrees of freedom in Jahn-Teller systems \cite{Kugel1982} and realized here for the first time in a truly {\em magnetic} variant. It is precisely this edge-sharing exchange coupling, which has been envisioned \cite{Jackeli2009} to lead to a realization of Kitaev couplings in the honeycomb iridates \NaIrO, \aLiIrO, and \RuCl\ and which has later also been found to be realized in the 3D hyper-honeycomb and strip-honeycomb materials \bLiIrO and \cLiIrO. 
This second scenario of edge-sharing exchange can be further expanded by considering octahedral geometries where neighboring IrO$_6$ octahedra do not share an edge, but have two parallel edges as illustrated in the right-most panel of Fig.~\ref{Fig:IridateExchangePaths}. Following the exact same line of arguments, one also arrives at a destructive interference for the isotropic Heisenberg exchange in this setting and the emergence of a bond-directional exchange. This latter scenario is realized in triangular Kitaev materials such as  Ba$_3$IrTi$_2$O$_9$.
\newline

In passing we note that recently also the microscopics of {\em face-sharing} ocahedra \cite{Streltsov2016} has  been discussed, which is relevant, for instance, to the quantum magnetism of Ba$_5$AlIr$_2$O$_{11}$ \cite{Terzic2015}.
\newline

Going beyond these symmetry-guided microscopic considerations and performing ab initio calculations \cite{Rau2014,Rau2016}, one finds that the generic Hamiltonian describing the interactions between $j=1/2$ spin-orbit entangled moments of spin-orbit assisted Mott insulators takes the general form
\begin{equation}
	H = -\sum_{\gamma \rm-bonds} J \,\, {\bf S}_i {\bf S}_j + K \,\, S_i^{\gamma} S_j^{\gamma} 
			+ \Gamma \left(  S_i^{\alpha} S_j^{\beta} + S_i^{\beta} S_j^{\alpha} \right) \,,
	\label{eq:HKG-model}
\end{equation}
where the sum runs over nearest-neighbor spins at sites coupled by a bond $\langle i,j \rangle$ along the $\gamma = (x,y,z)$ direction. The strength of the isotropic Heisenberg coupling is given by $J$, and the bond-directional couplings include (i) a Kitaev term of stength $K$ that couples the component $\gamma$ of the spins along a $\gamma$-bond and and (ii) a symmetric off-diagonal exchange $\Gamma$ that couples the two orthogonal spin components $\alpha, \beta \perp \gamma$ for a bond along the $\gamma = x,y,z$ direction.
The relative strength and coupling sign of the various couplings varies from material to material, but a common thread in all Kitaev materials is that the Kitaev coupling is the dominant exchange coupling, \ie $K > J, \Gamma$, with a similar ratio of the Kitaev to Heisenberg exchange of $|K/J| \approx 4$ for many $j=1/2$ Mott insulators (for a more detailed discussion, see below). 
The microscopic model \eqref{eq:HKG-model}, which is also referred to as the $JK\Gamma$ model, is often simplified to the Heisenberg-Kitaev model $(\Gamma = 0)$, which has first been conceptualized and studied in the context of the honeycomb iridates by Chaloupka, Jackeli, and Khaliullin \cite{Chaloupka2010}. We will return to this model after a discussion of the pure Kitaev model $(J, \Gamma = 0)$ in the next section.

\begin{figure}[h]
    \centering
     \includegraphics[width=0.4\hsize]{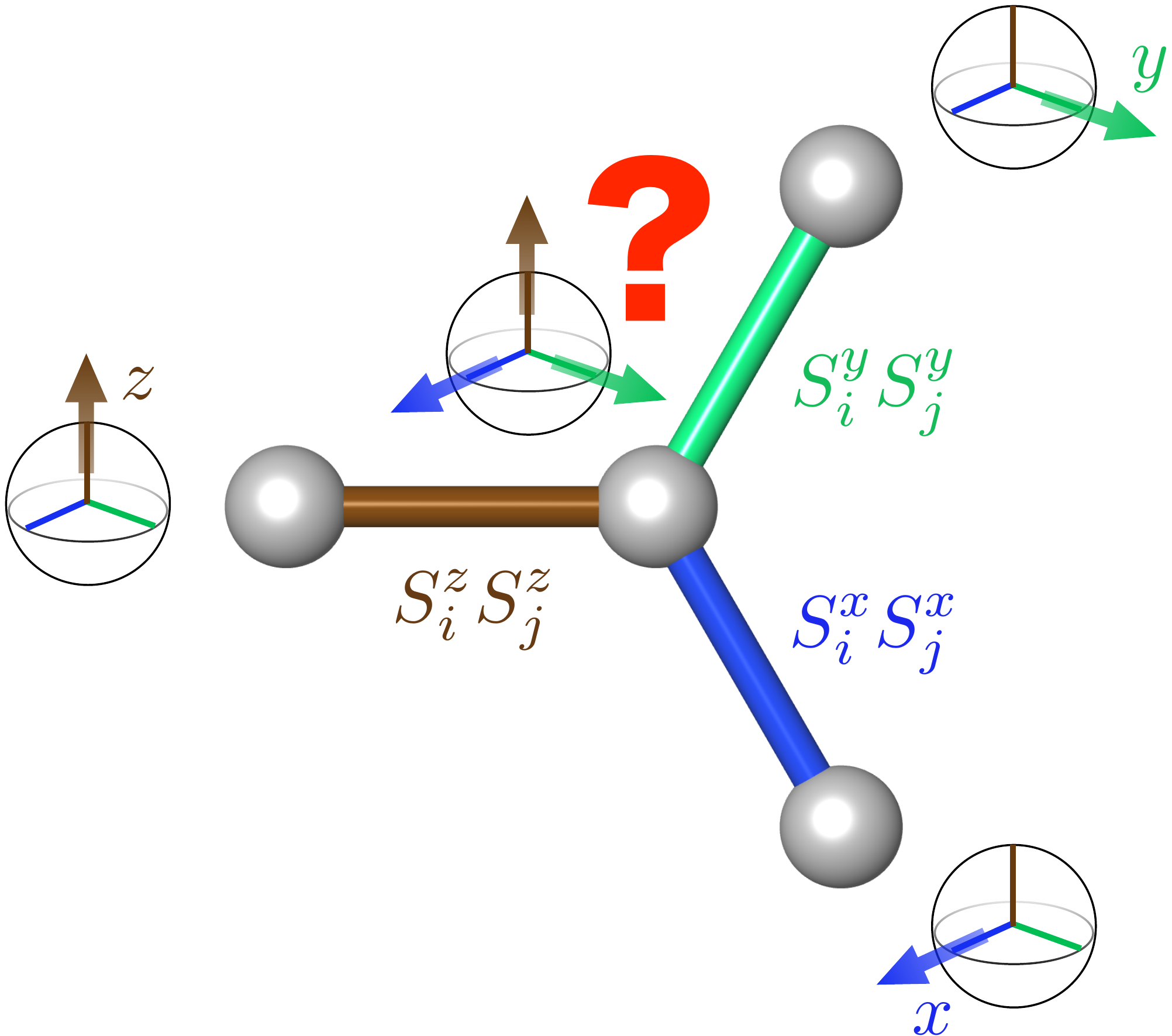}
     \caption{Exchange frustration arising from spin-orbit induced bond-directional interactions, \ie Ising-like couplings where the exchange 
     		easy axis depends on the spatial orientation of an exchange bond. Spins subject to these bond-directional interactions cannot 			simultaneously minimize all couplings, which holds both for quantum and classical moments.}
     \label{Fig:ExchangeFrustration}
\end{figure}

The effect of bond-directional interactions is a strong exchange frustration arising from the simple fact that these interactions cannot be simultaneously minimized, as illustrated for Kitaev-type couplings in Fig.~\ref{Fig:ExchangeFrustration}. Like geometric frustration, its more widely known counterpart that arises when the lattice geometry gives rise to constraints that cannot be simultaneously satisfied, the effect of exchange frustration is to inhibit magnetic ordering and give rise to a residual ground-state entropy. This is true already on the {\em classical} level. For instance, the classical Kitaev honeycomb model does not exhibit a finite-temperature phase transition \cite{Chandra2010,Sela2014}, but undergoes a thermal crossover to an extensively degenerate Coulomb phase \cite{Henley2010} at zero temperature.

\subsection{Kitaev model}

\begin{figure}[t]
    \centering
      \includegraphics[width=\hsize]{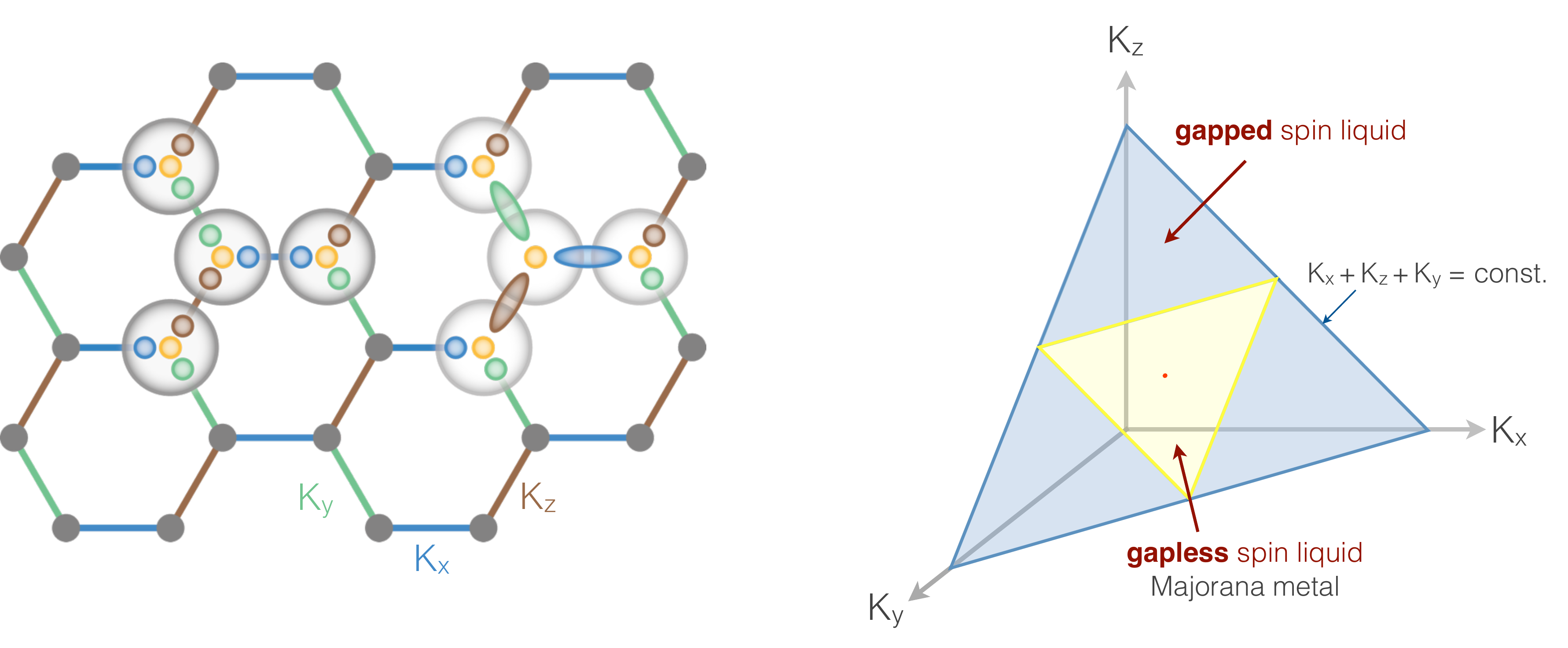}
     \caption{Left: The honeycomb Kitaev model with bond-directional couplings $K_x$, $K_y$ and $K_z$.
     		  The model can be analytically solved by introducing four flavors of Majorana fermions (indicated
		  by the yellow, blue, green and brown circles) and recombining them into a static \Z\ gauge field
		  (indicated by the blue, green and brown ovals) and a remaining itinerant Majorana fermion (yellow circle).
     		  Right: Phase diagram of the Kitaev model plotted for a plane $K_x + K_y + K_z = \text{const.}$
     		  If one of the three couplings dominates, the system forms a gapped spin liquid indicated by the blue shading.
		  Around the point of isotropic coupling strengths $K_x = K_y = K_z$ (indicated by the red dot) a gapless spin liquid
		  emerges, which can be best characterized as a (semi-)metal of the Majorana fermions.}
     \label{Fig:KitaevModel}
\end{figure}

The pure Kitaev model, which couples SU(2) spin-1/2 degrees of freedom with bond-directional interactions -- Ising-like couplings with a magnetic easy axis sensitive to the spatial orientation $\gamma$ of the bond (see Figs.~\ref{Fig:ExchangeFrustration} and \ref{Fig:KitaevModel} for an illustration)
\begin{equation}
  H_{\rm Kitaev} = -\sum_{\gamma \rm-bonds}  K_{\gamma} \,\, S_i^{\gamma} S_j^{\gamma}  \,,
  \label{eq:spinH}
\end{equation}
is of central interest in condensed matter physics (and beyond). For one, it is famously known for harboring both gapped and gapless quantum spin-liquid ground states. At the same time, it is one of the rare microscopic models that can be solved exactly, as demonstrated in seminal work \cite{Kitaev2006} by Alexei Kitaev in 2006. This analytical tractability allows one to precisely describe and track the fractionalization of the original spin-orbit entangled degrees of freedom ${\bf S}_i$ into a fermionic degree of freedom, a so-called Majorana fermion, and a \Z\ gauge field as illustrated in Fig.~\ref{Fig:KitaevModel}. The \Z\ gauge field turns out to be static, it orders at zero temperature and its elementary vison excitations are found to be massive. The Majorana fermions, on the other hand, remain itinerant and form a gapless state -- a {\em Majorana metal} -- around the point of equal coupling $K_x = K_y = K_z$. For the honeycomb lattice, this Majorana metal is a semi-metal with a Dirac cone dispersion (well known from the analogous calculation of free complex fermions for graphene-like electron systems). If one of the three couplings dominates, the system undergoes a phase transition (\eg for dominant $K_z$ coupling along the line $K_z = K_x + K_y$) into a gapped spin liquid. This latter state exhibits Abelian (\Z) topological order akin to the well-known toric code model \cite{Kitaev2003} and macroscopic entanglement. Applying a magnetic field along the 111-direction, \ie coupling the magnetic field to all three spin components, gaps out the gapless spin liquid into an even more exotic spin liquid with non-Abelian (Ising-type) topological order \cite{Kitaev2006}. The non-Abelian character of the latter is identical to that of a $p_x + i p_y$ superconductor \cite{Read2000}, the Moore-Read state \cite{Moore1991} proposed for the $\nu = 5/2$ fractional quantum Hall state, heterostructures of superconductors and topological band insulators \cite{Fu2008} or semicoductors \cite{Sau2010}, as well as that of a network  \cite{Alicea2011} of Majorana wires \cite{Lutchyn2010,Oreg2010} -- all physical systems, which have gathered considerable interest in the context of proposals for fault-taulerant topological quantum computation \cite{Kitaev2003,Nayak2008}. Despite this similarity, the search for Kitaev materials and a solid-state realization of the Kitaev model is probably less driven by a potential application in quantum computing technologies, but deeply inspired by the fundamental pursuit of (i) the synthesis of spin liquid materials, (ii) the experimental discovery of Majorana fermions, and (iii) a direct experimental probe of the underlying (\Z) gauge physics -- such experimental evidence for gauge physics in a condensed-matter context has long been lacking, despite theorists using the concept of \Z\ gauge theories in the classical statistical mechanics of nematics \cite{Lammert1993} and to capture the physics of fractionalization in quantum many-body systems  \cite{Read1991,Senthil2000,Wen2002} for decades.

\newpage

The conceptual understanding of the physics of the Kitaev model has been steadily growing since its initial description and analytical solution \cite{Kitaev2006}. This includes the fundamental role of vacancies \cite{Willans2010,Willans2011}, depletion \cite{Andrade2014}, and impurities \cite{Vojta2016}, disorder effects \cite{Willans2010,Lahtinen2014,Zschocke2015}, and more exotic phenomena such as the strain-induced formation of Landau levels  \cite{Rachel2016} or topological liquid nucleation \cite{Gils2009,Ludwig2011} arising form vortex-vortex interactions in the non-Abelian phase \cite{Lahtinen2012}. The emergence of $p$-wave superconductivity upon doping has been discussed \cite{Hyart2012,Kimchi2012,Trousselet2014,Halasz2014}.
With an eye towards a material realization of the (pure) Kitaev model several experimental signatures have been discussed including the dynamical response \cite{Knolle2014,Knolle2015}, whose distinct spectral gap can be probed via inelastic neutron scattering~(INS) or electron spin resonance~(ESR), the Raman response \cite{Knolle2014b,Perreault2016b}, the inelastic light scattering response \cite{Perreault2016}, which can be probed via resonant Raman scattering, as well as the resonant inelastic X-ray scattering~(RIXS) response \cite{Halasz2016}.

\begin{figure}[h]
    \centering
      \includegraphics[width=0.6\hsize]{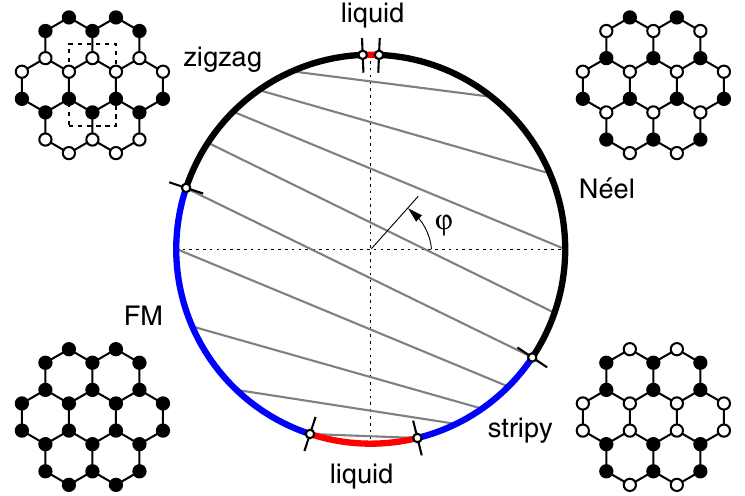}
     \caption{Phase diagram of the Heisenberg-Kitaev model, reproduced from Ref.~\cite{Chaloupka2013}.}
     \label{Fig:HKModel}
\end{figure}

Going beyond the pure Kitaev model, considerable attention has been devoted to the Heisenberg-Kitaev model. Its principal phase diagram \cite{Chaloupka2013} is shown in Fig.~\ref{Fig:HKModel} for a parametrization of the coupling strength $J=\cos \phi$, $K=\sin \phi$. Besides extended spin liquid phases around the pure Kitaev limit (with either ferro- or antiferromagnetic coupling) there are four magnetically ordered phases. This includes conventional ferromagnetic and N\'eel ordering around the pure Heisenberg limits, as well as a stripy and a zig-zag ordered phases. The origin of the latter can be readily understood by a duality \cite{Khaliullin2005} (sometimes referred to as Klein duality \cite{Kimchi2014}) of these phases to the two conventionally ordered ones. In particular, the Klein duality allows to map the left- and right-hand sides of the circular phase diagram of Fig.~\ref{Fig:HKModel} onto one another by the relation
\begin{equation}
  J \to -J \quad\quad\quad \text{and} \quad\quad\quad K \to 2J + K \,,
\end{equation}
and a concurrent four-sublattice rotation \cite{Khaliullin2005}. This duality thereby connects the ferromagnet to the stripy phase, and the N\'eel ordered state to the zig-zag ordered phase (as well as the spin liquid phases onto themselves).

The conceptual understanding of this extended model has been furthered by a discussion of its finite-temperature physics \cite{Reuther2011} and dynamical response \cite{Gohlke2017} along with the effects of an applied magnetic field \cite{Jiang2011}, lattice distortions \cite{Sela2014}, as well as the inclusion of further neighbor Heisenberg \cite{Kimchi2011} and Kitaev \cite{Rousochatzakis2015} interactions. The potentially multicritical point at the phase transition between the spin liquid and the magnetically ordered phases has been investigated both numerically \cite{Jiang2011} and analytically \cite{Schaffer2012}. The dynamical spin structure factor beyond the Kitaev limit has recently been discussed \cite{Song2016} as have been liquid-liquid transitions upon introducing an additional Ising coupling \cite{Nasu2016b}.
Extensions including charge fluctuations have also been discussed \cite{Laubach2017}.
Finally, a classical variant of the Heisenberg-Kitaev model has also been explored \cite{Price2012,Price2013,Sela2014}, including the formation of spin textures in a magnetic field \cite{Janssen2016,Chern2016}.


\section{Honeycomb Kitaev materials}

\begin{figure}[h]
    \centering
     \includegraphics[width=0.5\hsize]{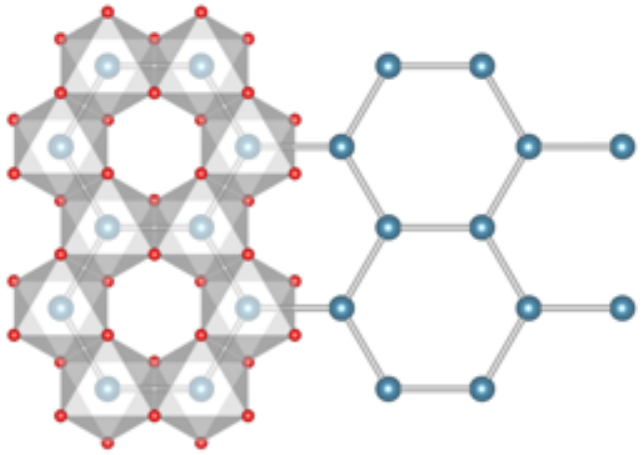}
     \caption{Crystal structure of the honeycomb Kitaev materials A$_2$IrO$_3$ such as \NaIrO\ and \aLiIrO.}
     \label{Fig:ChrystalStructureHoneycomb}
\end{figure}

The quest to synthesize and explore Kitaev materials has been kick-started in 2009 by the profound theoretical vision of Jackeli and Khaliullin \cite{Jackeli2009} that laid out in remarkably precise terms what the elementary ingredients (in its literal sense) for a successful material search strategy are. In particular, they not only explained the microscopic origin of Kitaev-type
bond-directional exchange interactions in certain (edge sharing) $4d^5$ and $5d^5$ transition metal compounds (as reviewed in the previous section). They also laid out a detailed proposal for a material realization of the honeycomb Kitaev model in the iridate  \aLiIrO\ and, more generally, iridates of the form A$_2$IrO$_3$ with a crystal structure as illustrated in  Fig.~\ref{Fig:ChrystalStructureHoneycomb} -- a proposal that was quickly refined \cite{Chaloupka2010} to also include \NaIrO\ which, at the time, had already been synthesized \cite{Takagi2009,Singh2010} and scrutinized as a potential topological insulator \cite{Shitade2009}.

\subsection{\NaIrO}

\NaIrO\ was independently synthesized to explore its potential for Kitaev physics in 2009/2010 by the groups of Takagi \cite{Takagi2009} and Gegenwart \cite{Singh2010}.
Since then samples have been grown in a number of labs around the world, including large single-crystals of diameters up to $10$~mm in the group of Cao \cite{Ye2012}. With samples readily available, numerous experimental probes have been taken that have revealed
that \NaIrO\ is indeed a Mott insulator with an insulating gap of $\Delta = 340$~meV (opening at around $300$~K) measured in optical transmission experiments \cite{Comin2012}. Fits of the magnetic susceptibility confirm the predominant $j=1/2$ nature of the local moments indicating magnetic moments of ~$1.79(2) \mu_B$ \cite{Singh2010,Singh2012}, rather close to the value of $1.74 \mu_B$ expected for spin 1/2 moments, while X-ray absorption spectroscopy \cite{Sohn2013} points to a small admixture of $j=1/2$ and $j=3/2$ states. Resonant inelastic x-ray scattering (RIXS) \cite{Gretarsson2013a,Gretarsson2013b} finds evidence for a small trigonal distortion of the IrO$_6$ octahedra resulting in a  crystal field splitting of the $j=3/2$ states of about $110$~meV, which, however, is considerably smaller than the typical strength of the spin-orbit coupling of the iridates $\lambda \approx 400-500$~meV \cite{Andlauer1976,Blazey1986}.

The system, however, does not exhibit the sought after spin liquid ground state, but is found to magnetically order at around $T_N = 15$~K, a temperature significantly below the Curie-Weiss temperature of $\Theta_{\rm CW} = -125$~K \cite{Singh2012}. The suppression of the ordering temperature with regard to the Curie-Weiss temperature indicates magnetic frustration which, quantified by the ratio $f = |\Theta_{\rm CW}| / T_N \approx 8$, is unusually high for a quantum magnet on a bipartite lattice. It has been rationalized \cite{Reuther2011,Singh2012} to arise from geometric frustration arising from next-nearest neighbor couplings within the elementary hexagons of the honeycomb structure and is not indicative of a close proximity to the Kitaev spin liquid. The exact form of the magnetic ordering was subsequently determined to be of zig-zag type by resonant x-ray magnetic scattering \cite{Liu2011} and inelastic neutron scattering \cite{Ye2012,Choi2012}. The microscopic origin of this zig-zag ordering has been scrutinized in light of the Heisenberg-Kitaev model \cite{Chaloupka2013} supplemented with next-nearest neighbor Heisenberg \cite{Singh2012} and Kitaev \cite{Sizyuk2015,Sizyuk2016} couplings along with a flurry of ab initio calculations \cite{Mazin2012,Foyevtsova2013,Mazin2013,Sohn2013} and, most recently, dynamical mean-field studies \cite{Yoshitake2016}. 

Probably the best experimental evidence that \NaIrO\ should indeed be considered to be the first Kitaev material to be synthesized comes from diffuse magnetic X-ray scattering \cite{Kim2015}, which has provided a direct experimental observation of bond-directional exchange and a dominant Kitaev coupling.

Several ideas have been put forward to bring \NaIrO\ closer to the Kitaev spin liquid regime, such as the making of thin films \cite{Yamaji2016} or heterostructures \cite{Winter2016}, but probably the most practical scheme has been to replace the Na atoms by the smaller Li atoms and instead explore the physics of \LiIrO.


\subsection{\aLiIrO}

The exploration of \LiIrO\ as candidate Kitaev material has indeed lent further impetus to the field. Probably the biggest surprise came when
its synthesis lead to the discovery of several polymorphs \cite{Singh2012,Takayama2015,Modic2014} which have been dubbed \aLiIrO, \bLiIrO, and \cLiIrO\ in the sequence of their discovery. Only \aLiIrO\ is isostructural to \NaIrO\ and exhibits iridium honeycomb layers (as depicted in Fig.~\ref{Fig:ChrystalStructureHoneycomb}), while the other two polymorphs, \bLiIrO\ and \cLiIrO, have been found to exhibit three-dimensional, tricoordinated networks of the iridium ions. We will discuss the latter in the context of three-dimensional Kitaev materials below.

The first high-quality samples of \aLiIrO\ have been synthesized by the Gegenwart group \cite{Singh2012} in 2012. Magnetic susceptibility fits reveal magnetic moment of ~$1.83(5) \mu_B$, a value slightly larger than expected for spin-1/2 degrees of freedom \cite{Singh2012}. Like its sister compound \NaIrO, \aLiIrO\ undergoes a magnetic ordering transition at $T_N = 15$~K with a  Curie-Weiss temperature of  $\Theta = -33$~K. Thin films of \LiIrO\ created by pulsed laser deposition on ZrO$_{\rm 2}$:Y (001) single crystalline substrates reveal \cite{Jenderka2015} a small optical gap of $~300$~meV indicative of a Mott gap of similar size as in its sister compound \NaIrO.

The further experimental analysis of \aLiIrO\ has been largely hampered by the fact that (in contrast to \NaIrO) it has remained extremely challenging to synthesize bulk single-crystals -- an essential ingredient \eg for neutron scattering studies, in particular to compensate for the large neutron absorption cross-section of the iridium ions. The early powder samples were found to exhibit single-crystalline ordering only on length scales smaller than $10~\mu$m. Recent experimental progress, however, has pushed this boundary to about a millimeter \cite{Freund2016}. 
These single crystals have been sufficiently large to allow resonant magnetic x-ray diffraction combined with powder magnetic neutron diffraction to reveal an incommensurate magnetic ordering where the magnetic moments in iridium honeycomb layers are {\em counter-rotating} on nearest-neighbor sites \cite{Williams2016}. 
Prior to its experimental observation a number of theoretical proposals \cite{Rau2014,Reuther2014,Chaloupka2015,Nishimoto2016} for spiral or other forms of incommensurate magnetic ordering in this honeycomb material have been put forward, none of which envisioned counter-rotating spirals. However, a very similar magnetic ordering is observed in the three-dimensional \bLiIrO\ and \cLiIrO\ polymorphs where subsequent theoretical analysis has lead to a unifying theory \cite{Kimchi2015,Lee2016} of counter-rotating spiral magnetism in all three compounds. The minimal model to explain the occurrence of counter-rotating spirals appears to be a Kitaev model supplemented with small Heisenberg and Ising interactions \cite{Williams2016} with theoretical estimates indicating that the Kitaev interaction might be up to a factor of 20 larger than the Heisenberg exchange \cite{Kimchi2015}.

Taking a step back, \aLiIrO\ remains much less explored than its sister compound \NaIrO\ despite its somewhat more unconventional magnetism, whose origin in a vastly enhanced Kitaev coupling puts \aLiIrO\ probably much closer to the sought-after Kitaev spin liquid than \NaIrO. This leads to the tempting observation that replacing the Na atoms by Li atoms indeed pushed the system closer to the Kitaev regime and that the next logical step would be to replace Li by an even smaller atom, \ie hydrogen. While a small step for a theorist to contemplate, this amounts to a big leap for material synthesis. But it is precisely this leap, which the Takagi group seems to have taken in the last few months by successfully synthesizing $\alpha$-H$_{3/4}$Li$_{1/4}$IrO$_3$  -- a hydrogen intercalated layered honeycomb iridate that shows a Curie-Weiss behavior consistent with a $j=1/2$ Mott insulator, no evidence of magnetic ordering, and no NMR line broadening down to the lowest temperatures \cite{Takagi2017}. These are prime indicators that this material might indeed realize a Kitaev spin liquid.


\subsection{\RuCl}

In parallel, another material has taken center stage in the search for Kitaev materials -- the $4d$ compound \RuCl.
The material consists of very weakly bounded layers of edge-sharing RuCl$_6$ octahedra (of almost perfect cubic symmetry with no trigonal distortions) with the central Ru$^{3+}$ ($4d^5$) ions forming an almost ideal honeycomb lattice. 
Originally thought to be a conventional semiconductor in early 1970's transport measurements \cite{Binotto1971}, spectroscopic measurements in the mid 1990's pointed towards the formation of a Mott insulator \cite{Pollini1996}. It was only in 2014 that the group of 
Young-June Kim realized that \RuCl\ is in fact a spin-orbit assisted $j=1/2$ Mott insulator \cite{Plumb2014}. Direct evidence for Mott physics in \RuCl\ comes from optical spectroscopy \cite{Plumb2014} measuring an optical gap of $200$~meV and angle-resolved photoemission spectroscopy (ARPES) \cite{Zhou2016} observing a charge gap of $1.2$~eV at temperatures of ~$200$~K. The strength of the spin-orbit coupling has been determined from optical spectroscopy \cite{Sandilands2016} to be $\lambda \approx 100$~meV and from neutron scattering experiments on polycrystalline samples \cite{Banerjee2016} to be $\lambda \approx 130$~meV, both estimates somewhat smaller than the atomic spin-orbit coupling $\lambda \approx 150$~meV for ruthenium \cite{Figgis1966}. While the strength of the spin-orbit coupling in this $4d$ compound is thus considerably smaller than for the heavier $5d$ iridates, it has been argued on the basis of ab initio calculations \cite{Plumb2014} that the {\em ratio} of the spin-orbit coupling and the electronic bandwidth is only slightly smaller than in the iridates and still suffices to induce the formation of spin-orbit entangled $j=1/2$ and $j=3/2$ bands. The formation of $j=1/2$ local moments is further supported by Curie-Weiss fits \cite{Kobayashi1992,Fletcher1967} of the magnetic susceptibility yielding a magnetic moment of $~2.2 \mu_B$ (somewhat above the expected value of $1.74 \mu_B$ for a spin-1/2) and by angle-resolved photoemission, x-ray photoemission, and electron energy loss spectroscopy \cite{Koitzsch2016,Sinn2016}. 
Scanning transmission electron and scanning tunneling microscopies \cite{Ziatdinov2016} on exfoliated/cleaved \RuCl\ samples report a Mott gap of $250$~meV in the measured density of states, slightly larger than the value obtained for bulk samples in optical spectroscopy, and a subtle charge ordering pattern originating from anisotropy in the charge distribution along Ru-Cl-Ru hopping pathways.

In exploring the magnetism of \RuCl,  susceptibility measurements on powder samples \cite{Kobayashi1992,Fletcher1967} indicate a Curie-Weiss temperature of around $23-40$~K consistent with single-crystal measurements \cite{Majumder2015} of the in-plane susceptibility yielding a value of $\Theta_{\rm CW} = 37$~K, while the out-of-plane susceptibility reveals a value of $\Theta_{\rm CW} = -150$~K.
At first two successive ordering transitions at $7$~K and $15$~K have  been reported, both in specific heat and susceptibility measurements on powder samples \cite{Majumder2015} as well as in single-crystal inelastic neutron scattering experiments  \cite{Banerjee2016}, with the higher transition later attributed to the presence of stacking faults in less pristine samples of \RuCl\ and entirely absent in high-quality single-crystals probed by X-ray diffraction \cite{Cao2016} and inelastic neutron scattering \cite{Banerjee2016b} experiments. Below $T_N = 7$~K pristine \RuCl\ exhibits zig-zag magnetic ordering (similar to that of \NaIrO) confirmed in various experimental probes including neutron \cite{Sears2015,Ritter2016} and X-ray \cite{Johnson2015,Cao2016} diffraction experiments, inelastic neutron scattering on polycrystalline \cite{Banerjee2016} and single-crystal \cite{Banerjee2016b} samples along with muon spin rotation measurements \cite{Lang2016}.
The emergence of zig-zag magnetic ordering in \RuCl\ is also consistent with ab initio calculations \cite{Kim2015c,Johnson2015,Kim2016c,Yadav2016} and microscopic calculations for the $H K \Gamma$ model \eqref{eq:HKG-model} \cite{Sizyuk2016} and variations thereof \cite{Chaloupka2016}. 
The strength of the Kitaev coupling in \RuCl\ has been estimated to be of the order of $100$~K in thermal conductivity measurements \cite{Hirobe2016}.
 
\begin{figure}[h]
    \centering
     \includegraphics[width=0.68\hsize]{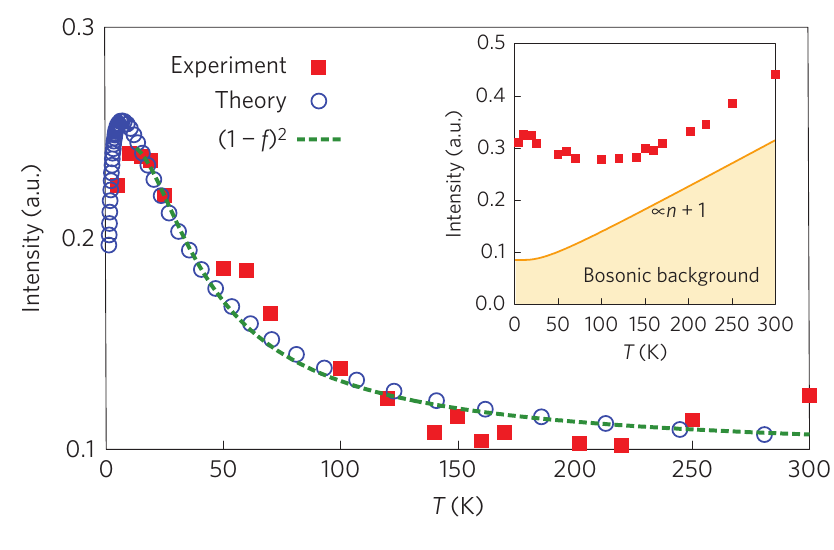}
     \caption{Interpretation of the Raman scattering data of \RuCl. 
     	            The inset shows the raw experimental Raman scattering intensities~\cite{Sandilands2015}, 
	            from which a (phononic) background contribution (indicated by the shaded area) is subtracted. 
	            The remaining contribution is shown in the main panel and compared to  
	            $(1-f)^2$, \ie the asymptotic two-fermion-scattering form where $f$ is the Fermi distribution function,
	            and numerical results~\cite{Nasu2016} from quantum Monte Carlo simulations (squares).
     		    Figure reproduced from Ref.~\cite{Nasu2016}.}
     \label{Fig:Raman-Data}
\end{figure}

What sets  \RuCl\ apart from the honeycomb iridates \NaIrO\ and \aLiIrO\ is that several unusual features {\em above} the magnetic ordering transition can be interpreted as arising from a close proximity to a Kitaev spin liquid. Here we want to highlight three such features found in Raman scattering and inelastic neutron scattering experiments of \RuCl.

We start with the Raman scattering experiments performed by Sandilands and collaborators \cite{Sandilands2015,Sandilands2016}.
Upon close inspection of the original Raman data \cite{Sandilands2015} plotted in the inset of Fig.~\ref{Fig:Raman-Data}, it has been argued by Nasu {\em et al.} \cite{Nasu2016} that the experimental data exhibits direct evidence of {\em fermionic} excitations across a broad energy and temperature range. This is an extraordinary observation, since the excitations of ordinary magnetic insulators, magnons and phonons, obey bosonic statistics. In the context of Kitaev spin liquids, however, fermionic excitations arise naturally, as one of the defining aspects of these spin liquids is the fractionalization of the original spin degrees of freedom into a \Z\ gauge field and a Majorana {\em fermion}. This should also hold for systems in close proximity to such a Kitaev spin liquid, which is precisely what Nasu {\em et al.} argue is the case for \RuCl\ \cite{Nasu2016}. The experimental evidence that is considered in their argument is the temperature dependence of the measured Raman intensity -- after subtraction of what is identified as a (bosonic) background contribution, which is attributed to phonons as it persists up to very high temperatures much larger than any magnetic scale. The remaining intensity distribution, plotted in the main panel of Fig.~\ref{Fig:Raman-Data}, is found to be very well approximated by $(1-f)^2$, \ie the asymptotic two-fermion scattering form where $f$ is the Fermi distribution function. An almost perfect fit is obtained when calculating the Raman intensity via quantum Monte Carlo techniques \cite{Nasu2014}. This is an intriguing result -- taken at face value, it is the first experimental indication for the emergence of (Majorana) fermion excitations in a bulk magnetic insulator. However, it should be noted that the interpretation of the experimental Raman data hinges -- in a crucial manner -- on the specific way that the (phononic) background contribution is identified and subtracted, an often subtle and not completely unambiguous procedure.

\begin{figure}[h]
    \centering
     \includegraphics[width=0.9\hsize]{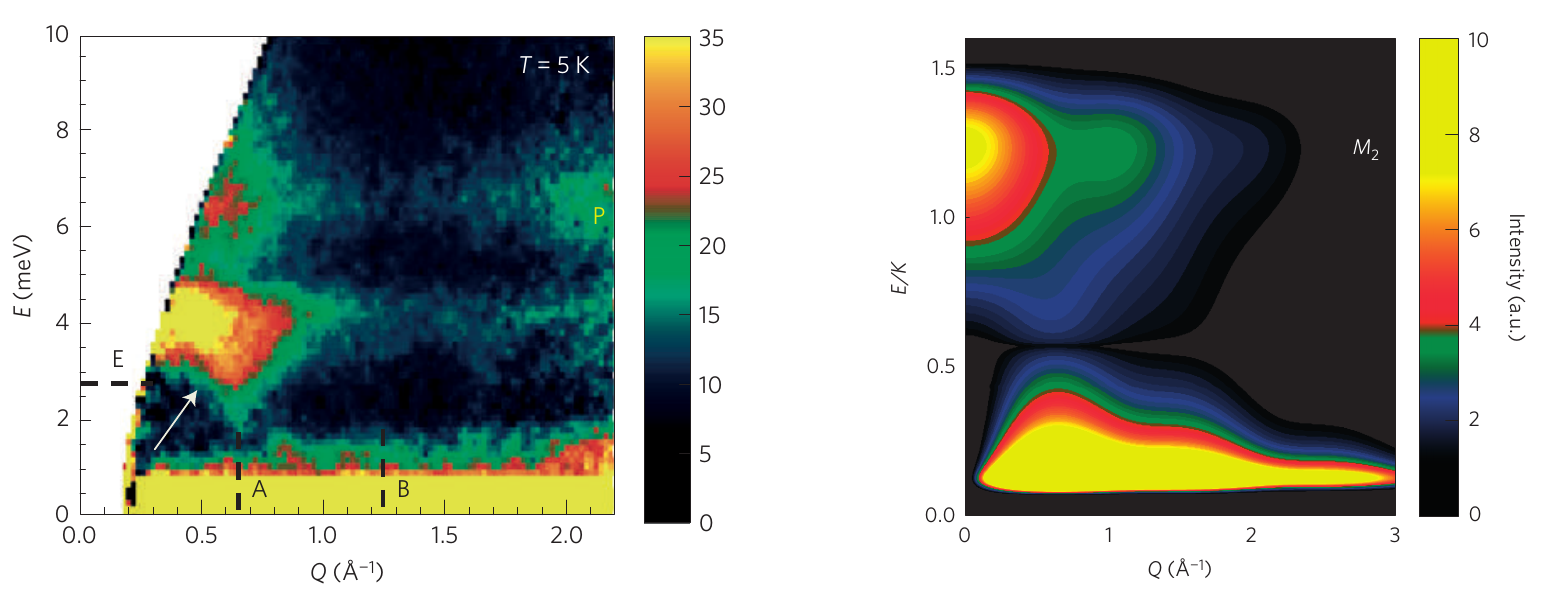}
     \caption{Left: False color plot of the inelastic neutron scattering data of \RuCl. 
     		  Two magnetic modes with band centers around $E=4$ and $6$~meV are identified. 
		  While the lower one is attributed to the concave spin wave dispersion on a zig-zag ordered background 
		  (indicated by the white arrow), 
		  the upper feature resembles the broad, non-dispersive high-energy response 	
		  expected of a Kitaev spin liquid.
		  Right: Analytical calculation of the dynamical response of the Kitaev spin liquid. 
		  Above a low-energy band a broad feature is found that is strongest at $Q=0$ and decreases with increasing $Q$.
     		  Figures reproduced from Ref.~\cite{Banerjee2016}.}
     \label{Fig:INS-Data-HighTemperature}
\end{figure}

\newpage

\begin{figure}[h]
    \centering
     \includegraphics[width=0.9\hsize]{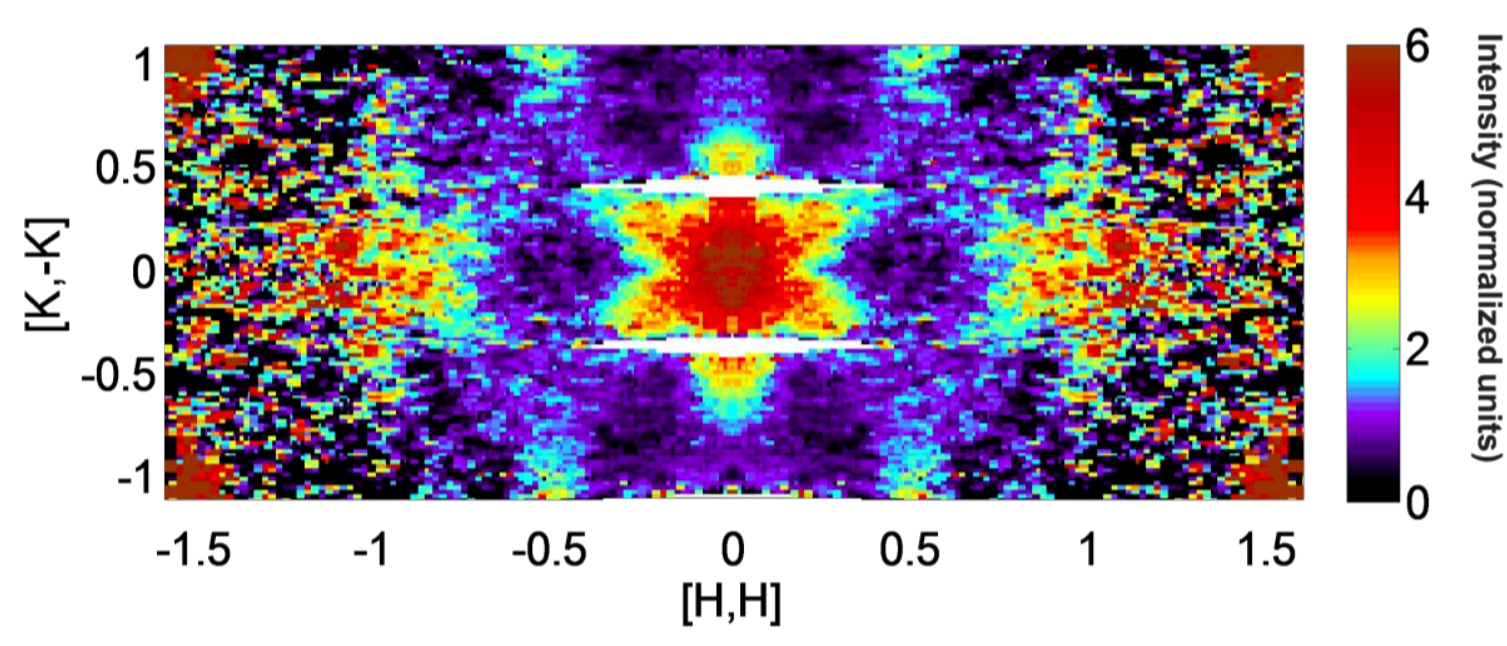}
     \caption{Extended zone picture of neutron scattering data of \RuCl\ integrated over the energy window $[4.5, 7.5]$~meV 
     		  and symmetrized along the $(H,H,0)$-direction
       		  taken at $T=10$~K, \ie well above the magnetic ordering transition at $T_N = 7$~K.
		  The star-like feature at the zone center arises from the interplay of spin wave and spin liquid physics 
		  in this temperature regime
		  and can be rationalized within the context of the Heisenberg-Kitaev model \cite{Gohlke2017}.
     		  Figure reproduced from Ref.~\cite{Banerjee2016b}.}
     \label{Fig:INS-Data-IntermediateTemperatureRegime}
\end{figure}

Two more experimental features strongly support the idea that \RuCl\ is indeed in close proximity to a Kitaev spin liquid, 
both of them are found in the magnetic scattering observed in inelastic neutron scattering experiments \cite{Banerjee2016,Banerjee2016b}. While at low energies the scattering is consistent with spin waves on a zig-zag 
ordered background, a broad scattering continuum is found at higher energies. It is this second magnetic mode that resembles the broad, non-dispersive high-energy response expected of a Kitaev spin liquid \cite{Banerjee2016}. A comparison of the experimental neutron scattering data with exact analytical calculations of the dynamical response of the Kitaev model are given in Fig.~\ref{Fig:INS-Data-HighTemperature}. 
At intermediate energy scales there are star-like features (reproduced in Fig.~\ref{Fig:INS-Data-IntermediateTemperatureRegime} \cite{Banerjee2016b}), which arise from the interplay of spin wave and spin liquid physics in this regime. 
While the dynamical response of the pure Kitaev model does not show this star-like feature, it can be explained by the small admixture of additional neighbor correlations as induced, for instance, by the inclusion of a Heisenberg exchange and experimentally observed in optical spectroscopy \cite{Sandilands2016b}. Indeed, recent numerical studies \cite{Gohlke2017} of the dynamical response of the Heisenberg-Kitaev model have quantitatively reproduced the star-like feature.

In total, the mounting experimental data on \RuCl\ support the proposal that, although \RuCl\ magnetically orders, it is in such close proximity to a Kitaev spin liquid that remnants of decisive spin-liquid features can still be probed {\em above} the magnetic ordering transition. This includes experimental evidence for fermionic quasiparticles in Raman scattering \cite{Sandilands2015,Nasu2016}, a broad magnetic continuum in inelastic neutron scattering \cite{Banerjee2016} and the star-like feature \cite{Banerjee2016b,Gohlke2017} above the magnetic ordering transition in inelastic neutron scattering.  
\newline

Let us finally mention that  also \RuCl\   has a polymorph, \bRuCl, in which face-sharing RuCl$_6$ octahedra are arranged in chains.
\bRuCl\ is found to show no magnetic ordering down to $5$~K \cite{Kobayashi1992} and awaits further experimental analysis.

%



\section{Triangular Kitaev materials}

Beyond the search for spin liquids, Kitaev materials also provide ample opportunity to study other unconventional forms of magnetism. 
A particularly interesting class might be materials in which $j=1/2$ moments form quasi-two-dimensional {\it triangular} lattice structures, such as the family of hexagonal perovskites Ba$_3$Ir$_x$Ti$_{3-x}$O$_9$, in particular the sister compounds Ba$_3$IrTi$_2$O$_9$ ($x=1$) \cite{Dey2012} and Ba$_3$TiIr$_2$O$_9$ ($x=2$) \cite{Sakamoto2006}. It has recently been argued \cite{Becker2015} that the microscopic description of the former is captured by a triangular Heisenberg-Kitaev model. The interplay of geometric and exchange frustration in this model leads to the formation of non-trivial spin textures \cite{Rousochatzakis2016,Becker2015}, such as the formation of a \Z-vortex crystal induced by the Kitaev couplings that destabilize the 120$^\circ$ order of the (antiferromagnetic) quantum Heisenberg model


\subsection{Ba$_3$Ir$_x$Ti$_{3-x}$O$_9$}

\begin{figure}[h]
    \centering
     \includegraphics[width=0.75\hsize]{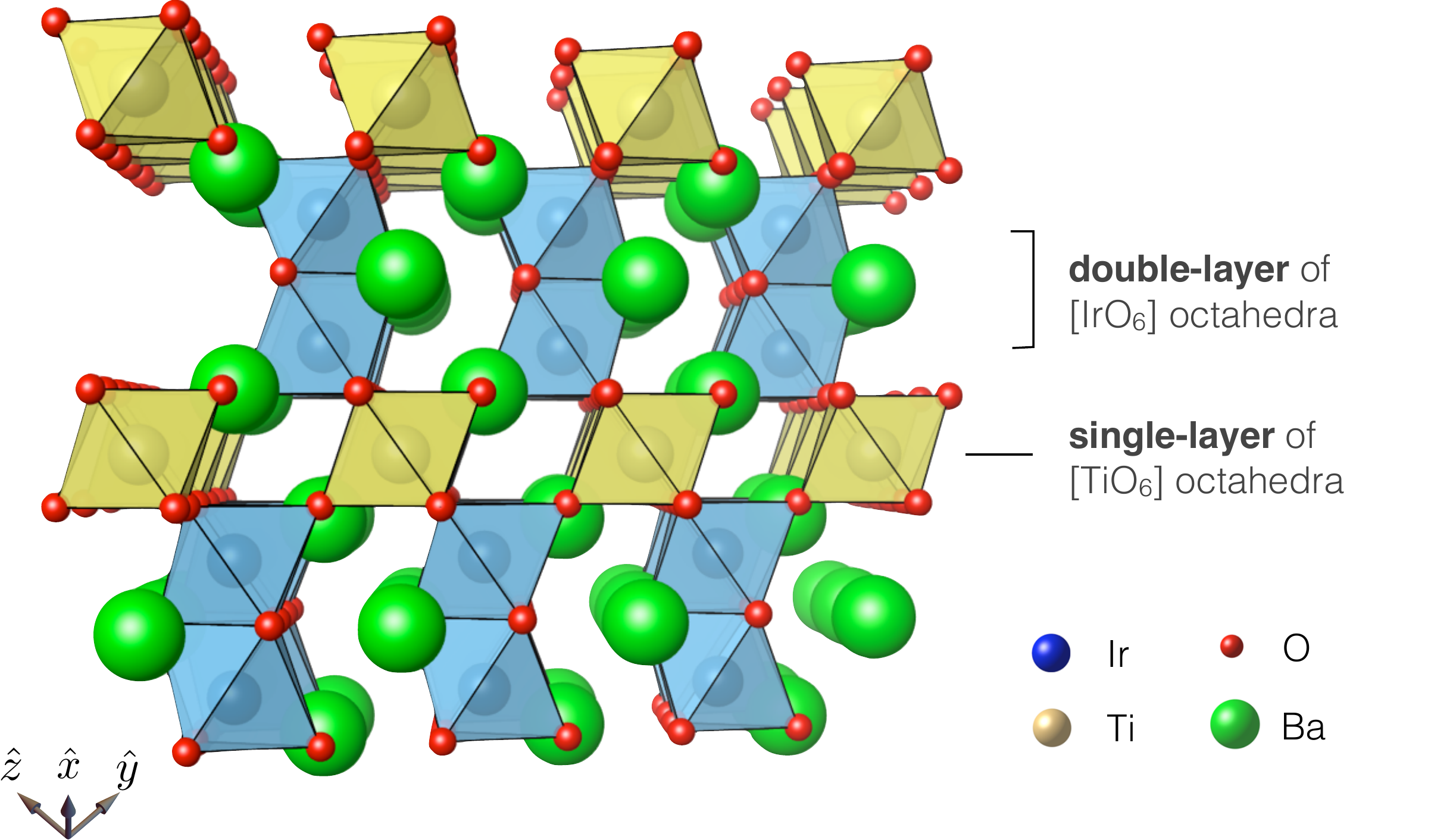}
     \caption{Crystal structure of the triangular Kitaev material Ba$_3$Ir$_2$TiO$_9$ \cite{Sakamoto2006}.}
     \label{Fig:Ba3Ir2TiO9}
\end{figure}

The 2012 synthesis of Ba$_3$IrTi$_2$O$_9$ ($x=1$) by the group of Mahajan \cite{Dey2012} marks the discovery of the 
first triangular $j=1/2$ Mott insulator. It is now recognized as the first representative of a family of hexagonal $j=1/2$ iridium perovskites of the form Ba$_3$Ir$_x$Ti$_{3-x}$O$_9$, of which various members with $1 \leq x \leq 2$ have been synthesized over the last years \cite{Kumar2016}.
The crystal structure of these perovskites -- which  are found to exhibit hexagonal space group symmetries P6$_3$mc ($x=1$) \cite{Dey2012} or P6$_3$/mmc ($x=2$) \cite{Sakamoto2006}, respectively -- is illustrated in Fig.~\ref{Fig:Ba3Ir2TiO9}.
Generically, the structure is composed of three layers of [Ir/TiO$_6$] octahedra, two of which form a double-layer of face-sharing Ir$_2$O$_9$ bioctahedra (indicated by the blue octahedra) and a single layer (indicated by the light brown octahedra). The occupation of the octahedral sites in the different layers with Ir$^{4+}$ and Ti$^{4+}$ ions is complicated by the fact that the two ions have rather similar radius. For Ba$_3$Ir$_2$TiO$_9$, it is found that the iridium ions fill all octahedral sites in the double-layer \cite{Sakamoto2006}, which are then well separated by a single layer of (non-magnetic) titanium ions. For Ba$_3$IrTi$_2$O$_9$ a more complicated picture emerges with the iridium ions still preferably occupying octahedral sites in one of two double-layers, but a significant amount of site inversion ($7\pm4\%$) with the titanium ions from the other double-layer has been reported \cite{Dey2012}. 

Magnetic susceptibility and heat capacity data \cite{Dey2012} for Ba$_3$IrTi$_2$O$_9$ show no magnetic ordering down to $0.35~K$ in spite of a strong antiferromagnetic coupling as evidenced by a large Curie-Weiss temperature $\Theta_{CW} \sim -130~K$. The effective moment is theoretically argued to be a spin-orbit entangled $j=1/2$ moment (on the basis of ab initio calculations \cite{Catuneanu2015}), despite the considerable suppression of the experimentally determined magnetic moment of ~$1.09~\mu_B$  \cite{Dey2012}.

\begin{figure}[h]
    \centering
     \includegraphics[width=0.95\hsize]{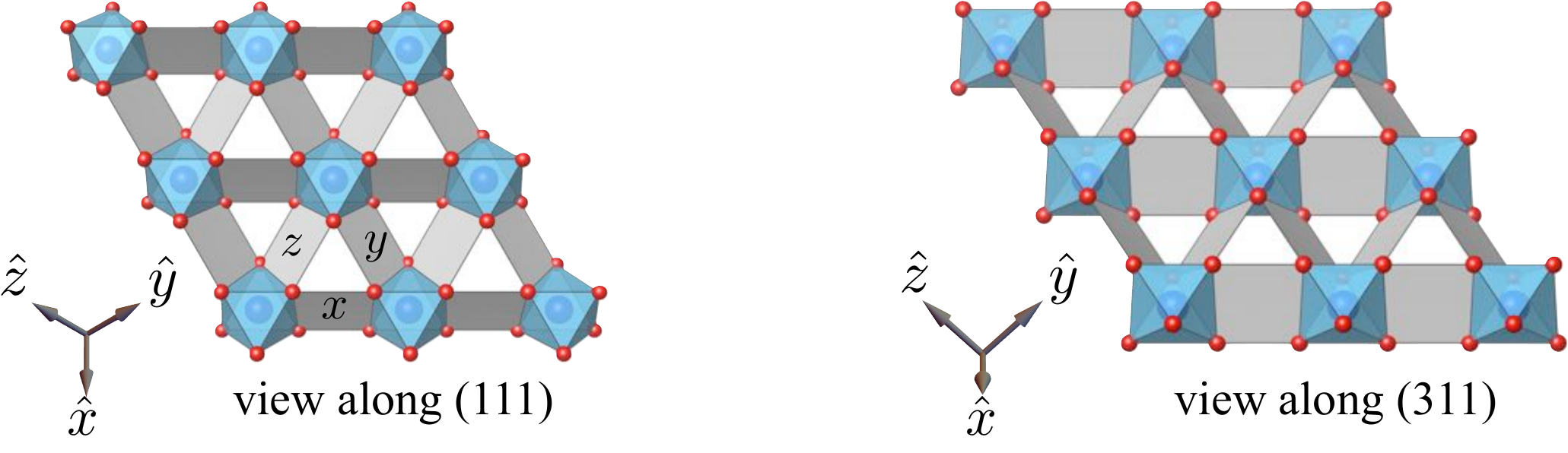}
     \caption{{Views of the triangular layer of IrO$_6$ octahedra from two different perspectives.}}
     \label{Fig:TriangularIridiumLayer}
\end{figure}

The unconventional magnetism of spin-orbit entangled $j=1/2$ moments on a triangular lattice again arises from the presence of bond-directional exchange couplings. On a microscopic level, it has been argued \cite{Becker2015} that the arrangement of the IrO$_6$ octahedra within the layers, depicted in Fig.~\ref{Fig:TriangularIridiumLayer}, fulfills the two necessary ingredients for Kitaev-type interactions. First, neighboring octahedra exhibit parallel edges, which gives rise to two separate exchange paths for every pair of iridium ions. As in the case of edge-sharing IrO$_6$ octahedra, this leads to a destructive interference and subsequent suppression of the isotropic Heisenberg exchange \cite{Khaliullin2005,Jackeli2009,Chaloupka2010}. Second, there are three distinct exchange paths for the three principal bond directions of the triangular lattice, with each cutting through {\em different} edges of the IrO$_6$ octahedra. This results in a distinct locking of the exchange easy axis \cite{Khaliullin2005,Jackeli2009,Chaloupka2010} along the three principal lattice directions as illustrated in Fig.~\ref{Fig:TriangularIridiumLayer}. Since the Ir layer is normal to the 111 direction (see the left panel in Fig.~\ref{Fig:TriangularIridiumLayer}), the strength of the bond-directional coupling is
equivalent in all three directions. Ultimately, this gives rise to the bond-directional exchange of a triangular Kitaev model. In total, these microscopic considerations lead to a triangular Heisenberg-Kitaev model as the most elementary description for the magnetism in Ba$_3$IrTi$_2$O$_9$. Ab initio calculations \cite{Catuneanu2015} complete this picture by arguing that in addition a symmetric off-diagonal exchange $\Gamma$ should be considered along with the possible emergence of a Dzyaloshinskii-Moriya (DM) exchange term arising from distortions in the oxygen octahedra, which break inversion symmetry about the Ir-Ir bond center.

\begin{figure}[h]
    \centering
     \includegraphics[width=0.93\hsize]{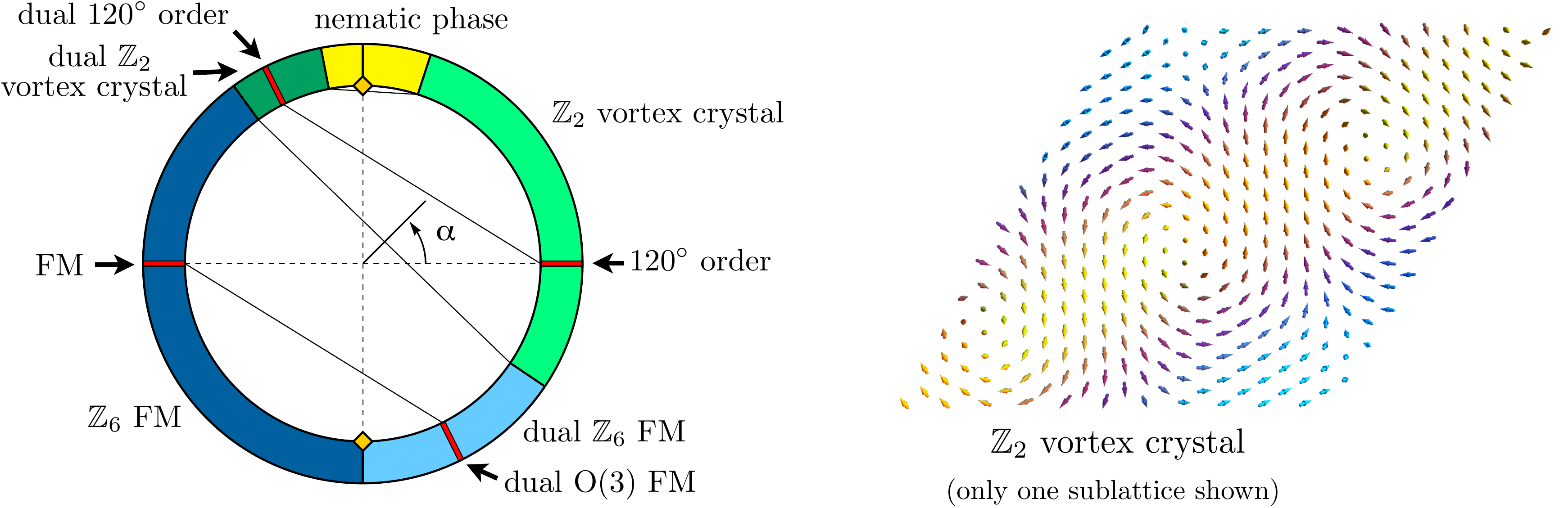}
     \caption{Phase diagram of the triangular Heisenberg-Kitaev model, reproduced from Ref.~\cite{Becker2015}.}
     \label{Fig:TriangularHK}
\end{figure}

The prevalent feature of the magnetism of the triangular Heisenberg-Kitaev model is the emergence of non-trivial spin textures \cite{Rousochatzakis2016,Becker2015}. The Kitaev exchange destabilizes the 120$^\circ$ order of the Heisenberg antiferromagnet and induces a lattice of \Z-vortices \cite{Rousochatzakis2016} whose spatial separation is inversely proportional to the strength of the Kitaev coupling (independent of its sign). The resulting phase diagram of the Heisenberg-Kitaev model has been explored both in its classical \cite{Rousochatzakis2016} and quantum \cite{Becker2015,Li2015,Kos2016,Shinjo2016} variants using a combination of analytical and numerical techniques. A summary is given in Fig.~\ref{Fig:TriangularHK}.

Finally, we note that other non-trivial spin textures beyond the \Z-vortex crystal can be stabilized by spin-orbit coupling effects. For instance, a Dzyaloshinskii-Moriya (DM) exchange also destabilizes the 120$^\circ$ order of the Heisenberg antiferromagnet and instead favors the formation of a skyrmion crystal in the presence of a magnetic field \cite{Okubo2012,Rosales2015}. 


Returning to the family of Ba$_3$Ir$_x$Ti$_{3-x}$O$_9$ materials, it would be most intriguing to experimentally establish the emergence
of non-trivial spin textures in Ba$_3$IrTi$_2$O$_9$. This, however, requires high-quality single crystals (without the aforementioned Ir/Ti site inversion) that would allow for inelastic neutron scattering experiments. The magnetism of the sister compound Ba$_3$TiIr$_2$O$_9$ is likely dominated by the formation of dimers in the Iridium double-layer and is currently being explored both theoretically and experimentally.


\subsection{Other materials}

Recently, a new family of hexagonal $j=1/2$ iridium perovskites of the form Ba$_3$$M$Ir$_2$O$_9$ with $M = (\text{Sc, Y})$ has been experimentally explored \cite{Dey2013,Dey2014}. In contrast to the Ba$_3$Ir$_x$Ti$_{3-x}$O$_9$ compounds, the (average) iridium valence here is Ir$^{4.5+}$. This possibly leads to a scenario wherein the double-layer one has one effective spin-orbit entangled $j=1/2$ moment per face-sharing Ir$_2$O$_9$ bioctahedra. As argued above, the latter are coupled via three distinct parallel (bi)octahedral edges and as such the effective $j=1/2$ moments are likely subject to a bond-directional exchange. While both Ba$_3$ScIr$_2$O$_9$ and Ba$_3$YIr$_2$O$_9$ have been reported to exhibit magnetic ordering at around $10$~K (Sc) and $4.5$~K (Y), respectively, the closely related Ir$^{4.5+}$-compound Ba$_3$InIr$_2$O$_9$ \cite{Sakamoto2006} does not exhibit any sign of magnetic ordering down to $250$~mK \cite{Dey2016} and, thus, is a potential $j=1/2$ spin liquid candidate system.

Another related spin liquid candidate material is the recently synthesized ruthenate Ba$_3$ZnRu$_2$O$_9$ \cite{Terasaki2017}, in which a hexagonal lattice of Ru$^{5+}$ dimers (with $S=3/2$) forms in the double-layer of face-sharing Ru$_2$O$_9$ bioctahedra. The absence of long-range magnetic order down to $37$~mK along with a linear specific heat \cite{Terasaki2017} indicate the possible formation of a spin liquid in this material, which would be remarkable given the rather large effective magnetic moment of $S=3/2$.


\section{Three-dimensional Kitaev materials}

The exploration of three-dimensional Kitaev materials was kick-started with the independent, but almost concurrent synthesis of two \LiIrO\ polymorphs in 2013 -- \bLiIrO\ in Takagi's group \cite{Takayama2015} and \cLiIrO\ in the group of Analytis \cite{Modic2014}. Both polymorphs realize truly three-dimensional, but still tricoordinated sublattices of the iridium $5d^5$ ions, dubbed the hyper-honeycomb and stripy-honeycomb, respectively. As such, these compounds are candidate materials for the realization of three-dimensional Kitaev physics, which we will briefly review in the following before returning to the materials.


\subsection{Conceptual overview}

On a conceptual level, it has been realized early on that the original honeycomb Kitaev model \cite{Kitaev2006} can be generalized to other lattice structures and retain its analytical tractability if one preserves one essential feature of the honeycomb lattice -- the {\em tricoordination} of its vertices. This idea has been exploited, for instance, by Yao and Kivelson \cite{Yao2007} in designing a decorated honeycomb lattice for which the Kitaev model exhibits a {\em chiral spin liquid} -- a distinct gapped spin liquid originally proposed by Kalmeyer and Laughlin in 1987 \cite{Kalmeyer1987} as a bosonic analog of the fractional quantum Hall effect\footnote{Microscopic realizations of this long sought-after spin liquid in SU(2) invariant quantum magnets have more recently been established in certain kagome systems  \cite{Bauer2014,He2014,Gong2014,Gong2015}.}.
The potential for {\em three-dimensional} generalizations of the Kitaev model\footnote{
Alternative approaches to generalize the Kitaev model to three spatial dimensions also include higher spin generalizations such as a spin-3/2 Kitaev-type model on the the diamond lattice \cite{Ryu2009} as well as initial work \cite{Si2007,Si2008} identifying spin-1/2 Kitaev models on certain three-dimensional, tricoordinated helix lattices.
}
 has first been explored by Mandal and Surendran \cite{Mandal2009} by considering a site-depleted cubic lattice, a tricoordinated lattice structure which turns out to be isomorphic to the hyper-honeycomb lattice later identified in the context of the iridium sublattice in \bLiIrO. This first three-dimensional generalization of a Kitaev model illustrates that much of the physics of the two-dimensional honeycomb Kitaev models carries over to higher dimensions: The original spin degrees of freedom again fractionalize into Majorana fermions coupled to a \Z\ gauge field. As in the two-dimensional case, the gauge field remains static, which again allows to analytically track the model by (i) solving a basically classical problem identifying the ground state of the \Z\ gauge field and (ii) subsequently diagonalizing a free fermion Hamiltonian describing the physics of the remaining itinerant Majorana fermions (coupled to a fixed gauge field configuration). The result is a gapless spin liquid whose nodal structure is no longer a pair of Dirac cones (as for the two-dimensional honeycomb model), but in fact a {\em line} of Dirac cones \cite{Mandal2009}.
This result has foreshadowed a more systematic understanding of three-dimensional Kitaev models \cite{OBrien2016} obtained from a systematic classification of the gapless spin liquids in Kitaev models on the most elementary tricoordinated lattices in three spatial dimensions. 

\begin{figure}[h]
    \centering
     \includegraphics[width=0.95\hsize]{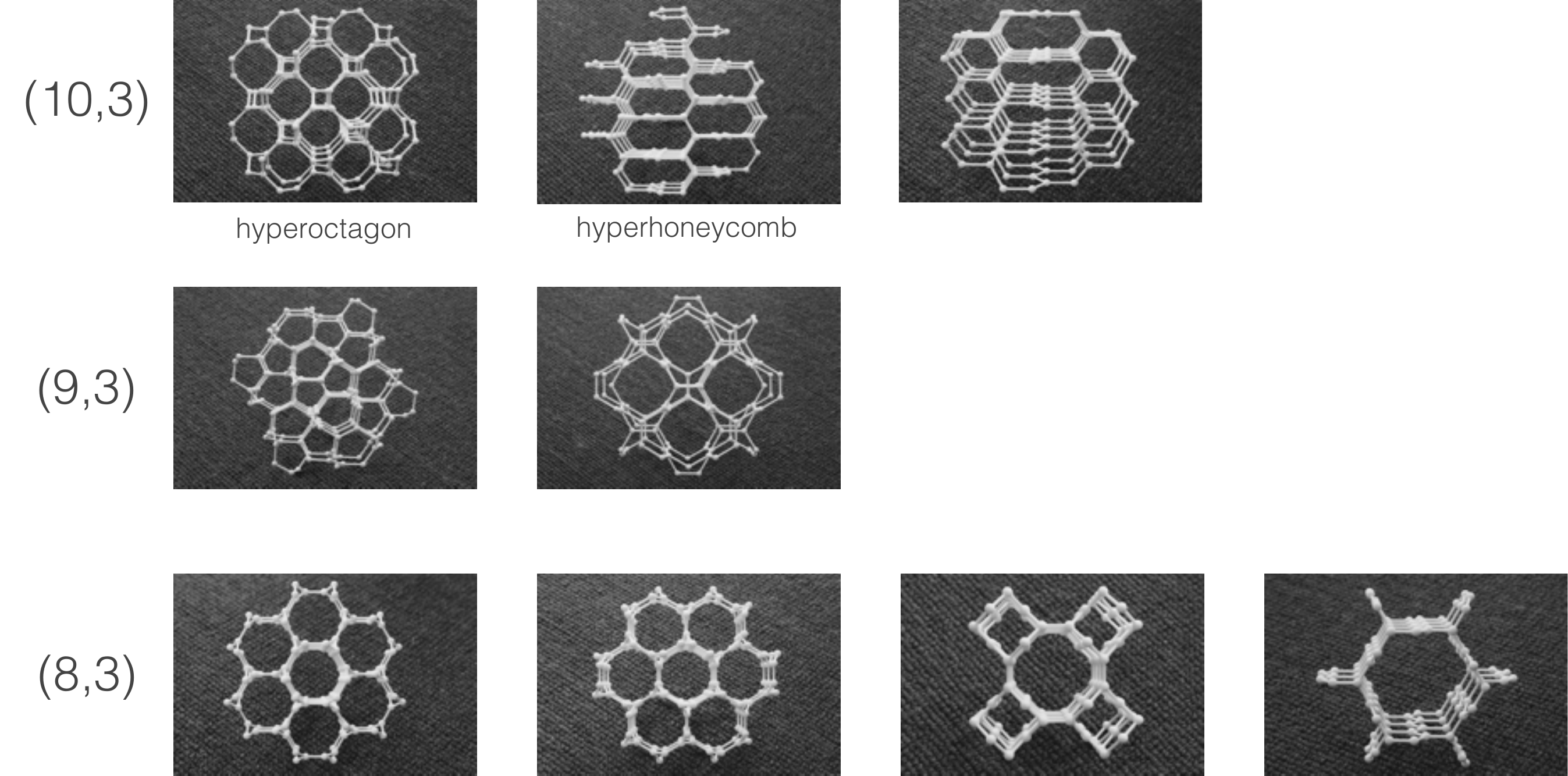}
     \caption{Illustration of the elementary tricoordinated lattices by photographs of 3D printed models.
     		  Further information on these lattices is provided in Table \ref{Tab:3Dlattices}.
     		}
     \label{Fig:3Dlattices}
\end{figure}

\begin{table}[t]
    \centering
     \includegraphics[width=0.65\hsize]{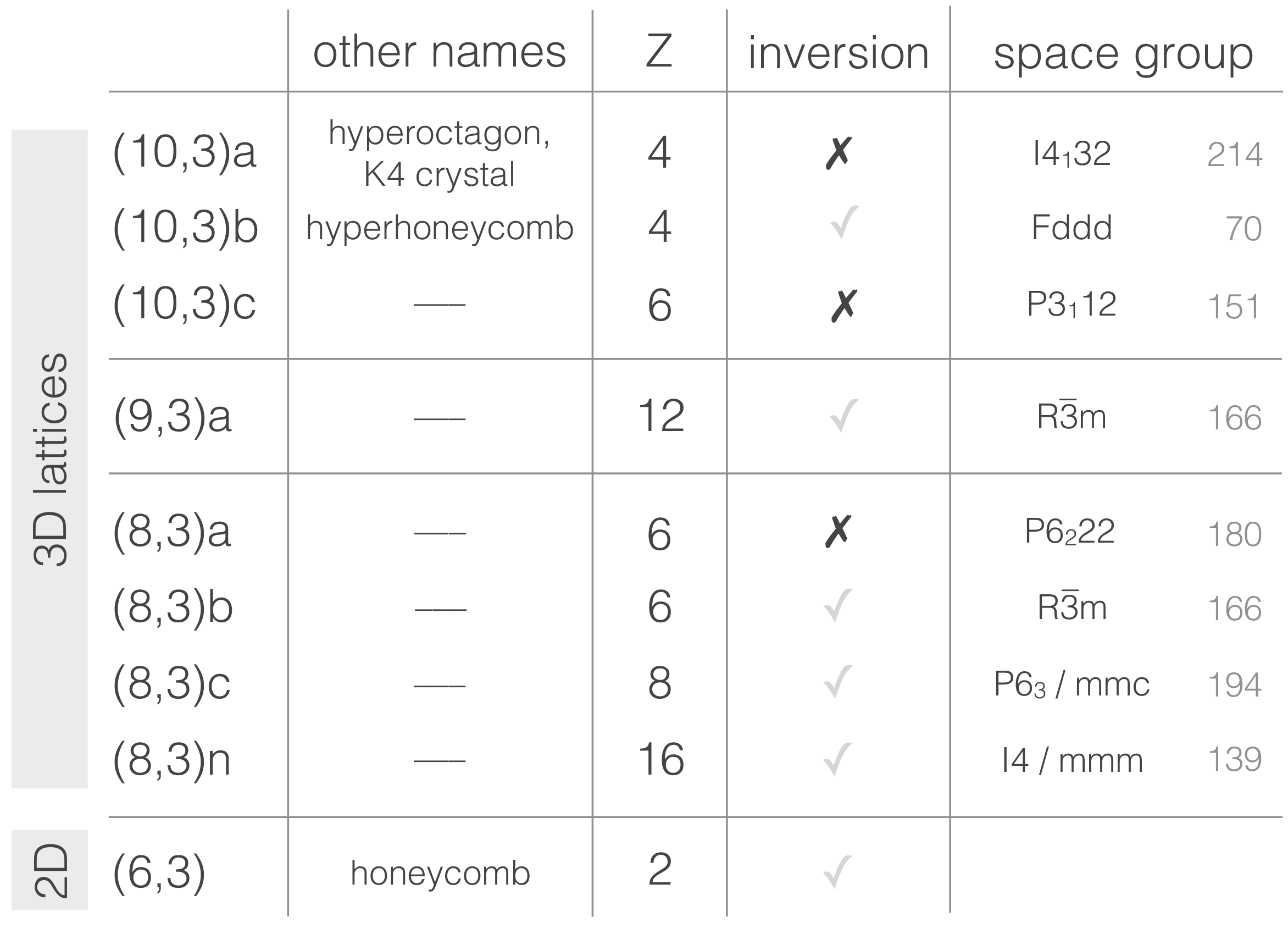}
  \caption{Overview of tricoordinated lattices in two and three spatial dimensions. 
  		Each lattice is described by its Schl\"afli symbol $(p,c)$ identifying the length of the elementary loops (or polygonality) $p$
		and tricoordination $(c=3)$. Along with alternative names used in the literature some basic lattice information is provided 
		including the number of sites $Z$ in the unit cell, an indication whether the lattice exhibits inversion symmetry, and its
		space group description.
  }
  \label{Tab:3Dlattices}
\end{table}

In fact, tricoordinated lattices in three spatial dimensions have been exhaustively classified by Wells in the 1970's \cite{Wells1977}.
The most elementary ones are lattice structures where all plaquettes (\ie shortest loops within the lattice) have the {\em same} length (which is also referred to as the polygonality $p$ of the lattice). While for two spatial dimensions there is only one such elementary tricoordinated lattice -- the honeycomb lattice of polygonality 6, there are multiple elementary lattices of higher polygonality 8, 9, 10 and higher, which are all three-dimensional.  An overview of these elementary tricoordinated lattices is given in Table \ref{Tab:3Dlattices}, where each lattice is labeled by its Schl\"afli symbol $(p,3)$ and a letter that simply enumerates the lattices for a given Schl\"afli symbol (the 3 indicates its tricoordination). Fig.~\ref{Fig:3Dlattices} illustrates these lattice structures via photographs taken from 3D printed models. 
For each of these lattice structures there is precisely one assignment of bond-directional Kitaev-type couplings that respects all symmetries of the lattice (up to a trivial permutation among the couplings). The so-defined Kitaev models can all be solved analytically\footnote{
In all fairness, it should be noted that this is not entirely true in a rigorous sense. While for the two-dimensional honeycomb model the ground state of the of the \Z\ gauge field can be readily inferred from a theorem of Lieb \cite{Lieb1994}, this is not generically true for the three-dimensional lattices. In fact, only one of the lattices, (8,3)b, strictly allows for the application of the theorem, while for all other lattices one has to resort to alternative means such as numerics. Thus, in a more rigorous sense, only the Kitaev model on lattice (8,3)b can be solved exactly.
}
and for all but one lattice, (8,3)n, the ground state is found to be a gapless spin liquid. This state is best described as a {\em Majorana metal} formed by the itinerant Majorana fermions, while the elementary excitations of the (static) \Z\ gauge field -- the visons -- are gapped. The precise nature of these Majorana (semi)metals turns out to depend on the underlying lattice geometry and 
can be captured, \eg, by its nodal manifold \cite{OBrien2016}. As described for the hyper-honeycomb (corresponding to the (10,3)b lattice in the above classification) considered by Mandal and Surendran \cite{Mandal2009}, this nodal manifold can be a nodal line.
Other possibilities for the nodal structure include the formation of Majorana Fermi surfaces \cite{Hermanns2014} or even topologically protected\footnote{
The topological nature of the bulk band structure is also reflected in gapless surface states \cite{OBrien2015,Schaffer2015,OBrien2016} such as Fermi arcs for the Weyl spin liquids.} 
Weyl nodes \cite{OBrien2015}. 
Table~\ref{Tab:majorana_metals} provides a complete classification of the Majorana metals of all elementary three-dimensional lattices,
which notably contains multiple examples each for the emergence of Fermi surfaces, nodal lines, or Weyl nodes for the various lattice structures. For a given lattice, the precise nature of these Majorana metals can in fact be deduced from an elementary symmetry analysis of the projective time-reversal and inversion symmetries as discussed in detail in Ref.~\cite{OBrien2016}.
Going beyond these most elementary lattices one might consider, for instance, higher harmonics of a given elementary lattice by systematically enlarging some plaquettes while simultaneously shortening others, see Ref.~\cite{Modic2014} for further details and showcasing that the stripy-honeycomb lattice is in fact a higher harmonic of the elementary (10,3)b hyper-honeycomb lattice.  The classification scheme described above covers all these higher harmonics as well, as the nature of the emergent Majorana metal remains unchanged going from an elementary lattice to any of its higher harmonics\footnote{This excludes the limiting case of the ``infinite harmonic'', which in fact is a two-dimensional lattice, \eg the square-octagon lattice for lattice (10,3)a or the honeycomb lattice for the hyper-honeycomb lattice (10,3)b.}.

\begin{table}[t]
    \centering
     \includegraphics[width=0.75\hsize]{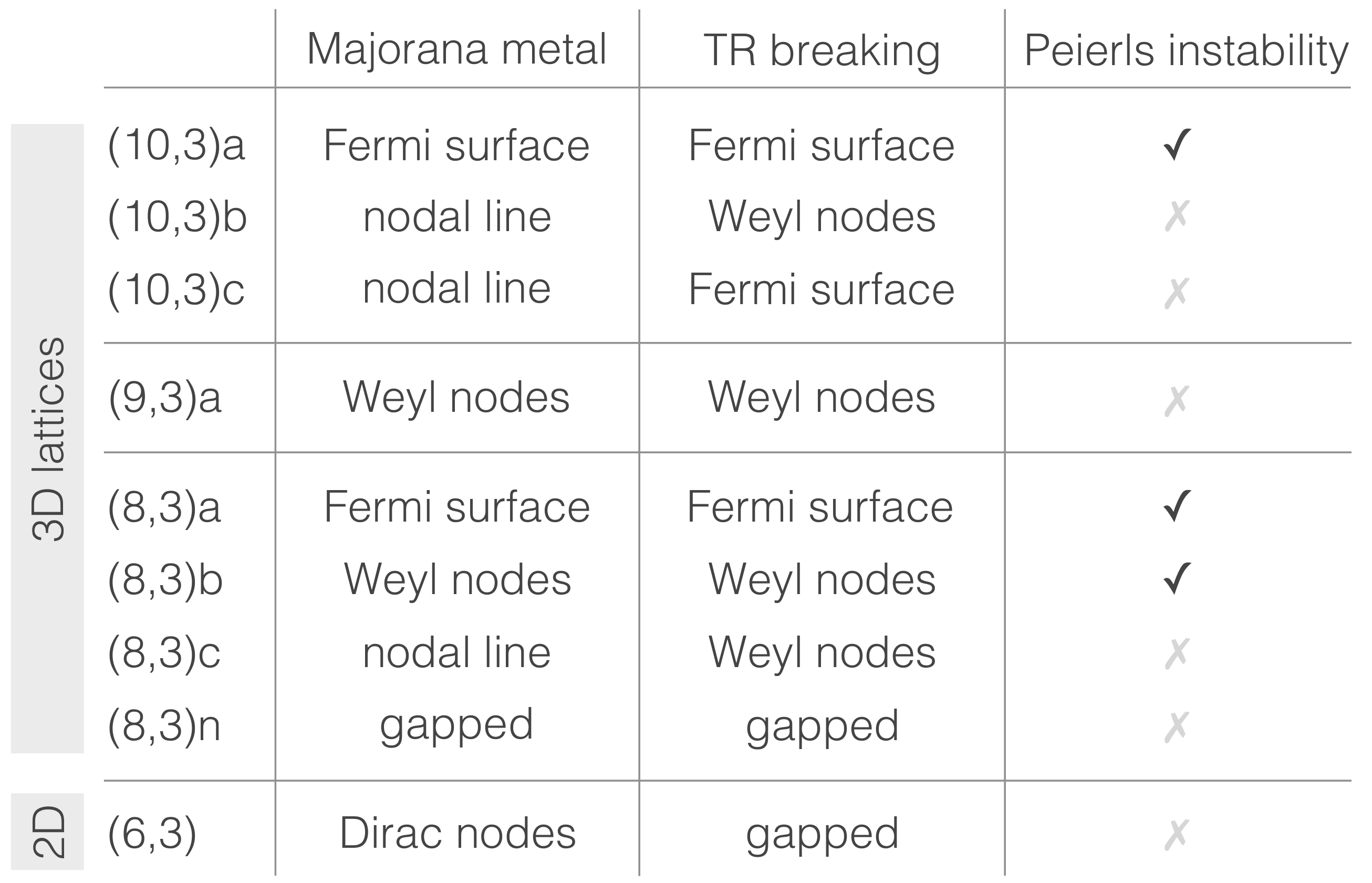}
  \caption{Overview of Majorana metals characterizing the gapless spin liquids in three-dimensional Kitaev models. 
  		Depending on the underlying lattice geometry, different Majorana (semi)metals are formed by the itinerant
		Majorana fermions, which are characterized here via their nodal structures. 
		Results for the pure Kitaev model \eqref{eq:spinH} are given in the second column, while the third
		column provides information on how the metallic nature (and its nodal structure) changes when 
		time-reversal symmetry (TRS) is broken, \eg by augmenting the Kitaev model by a magnetic field term 
		(pointing along the 111-direction). The last column indicates whether the metallic state is susceptible to 
		a spin-Peierls instability \cite{Hermanns2015}.
	 Table adapted from Ref.~\cite{OBrien2016}.
  }
  \label{Tab:majorana_metals}
\end{table}

What does change the nature of the Majorana metal for a given lattice though is the application of a magnetic field or, more generally, the breaking of time-reversal symmetry. For the two-dimensional honeycomb lattice a particularly interesting scenario occurs -- the Dirac spin liquid gaps out into a massive non-Abelian topological phase when applying a magnetic field along the 111 direction, \ie a magnetic field that couples to all three spin components
\begin{equation}
  H= -\sum_{\gamma \rm-bonds}  K_{\gamma} \,\, S_i^{\gamma} S_j^{\gamma} - \sum_j  {\bf h} \cdot {\bf S}_j \,,
  \label{eq:KitaevMagneticField}
\end{equation}
with ${\bf h} = h\, (1,1,1)^t$. Keeping in mind that the Kitaev model reduces to a free (Majorana) fermion model, this scenario can be rationalized by considering the classification of topological insulators \cite{Schnyder2008,Kitaev2009,Ryu2010} rooted in the symmetry classification of free-fermion systems \cite{Altland1997}. While the unperturbed Kitaev model resides in symmetry class $BDI$,  the model with broken time-reversal symmetry is in symmetry class $D$ for which the classification scheme of topological insulators indeed points to the possibility of a topologically non-trivial band insulator in two spatial dimensions. For three spatial dimensions, in contrast, this scenario of driving the (non-interacting) Majorana metal into a topologically non-trivial gapped phase is not possible. Instead, it is found that in the presence of a time-reversal symmetry breaking field (again pointing along the 111-direction), the three-dimensional Kitaev models remain gapless, but the nature of the Majorana metal might change. This is, for instance, the case for the nodal line metals, which in the presence of a magnetic field turn either into Weyl semimetals or Weyl metals, \ie the energy spectrum acquires Weyl nodes in both cases sitting right at the Fermi energy in the first case and above/below the Fermi energy in the second case \cite{OBrien2016}. The effect of  time-reversal symmetry breaking  for all Majorana metals is provided in the third column of Table \ref{Tab:majorana_metals}.

\begin{figure}[t]
    \centering
     \includegraphics[width=0.75\hsize]{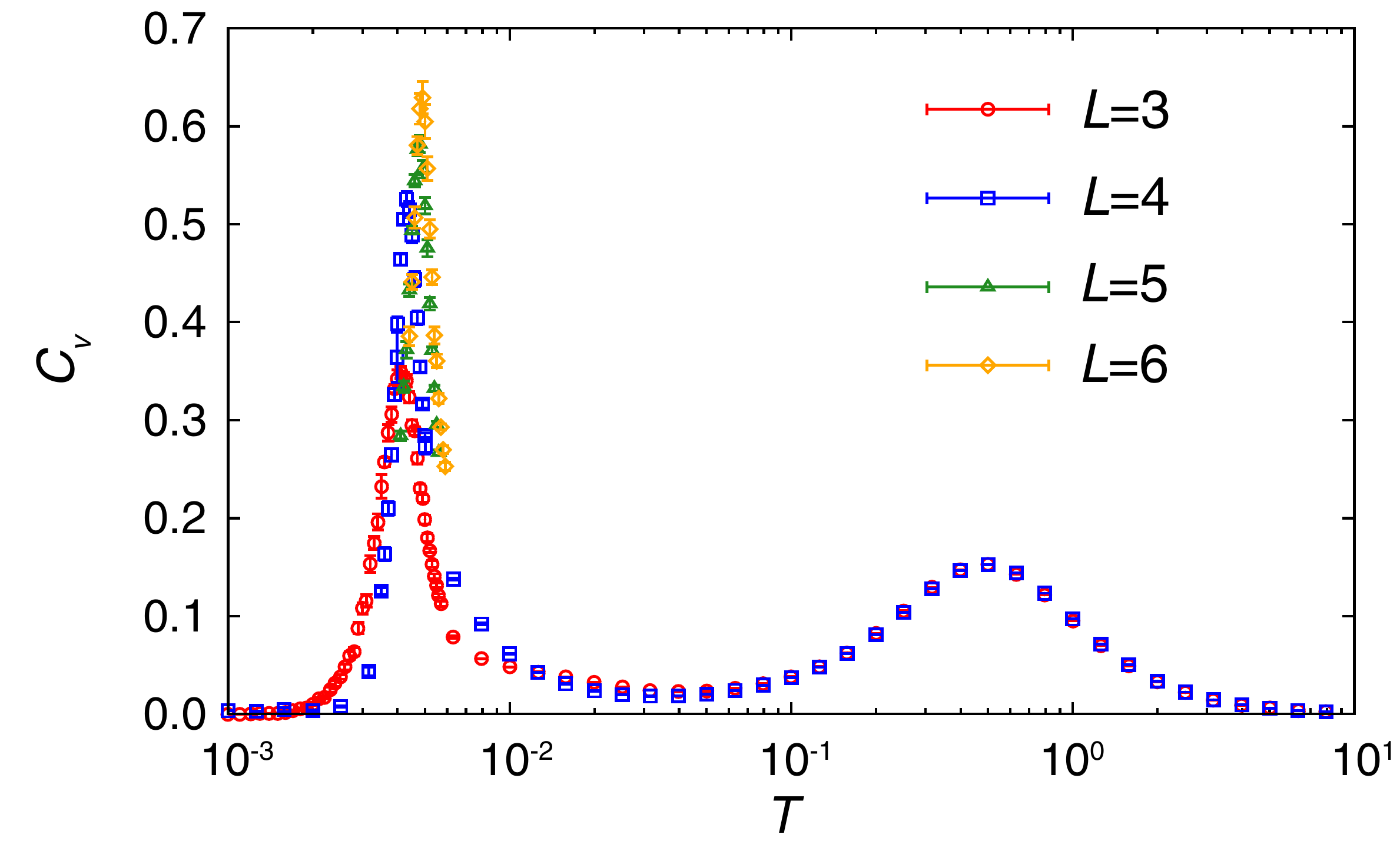}
     \caption{Signatures of spin fractionalization and a subsequent \Z\ gauge transition in the specific heat 
     		calculated for the hyper-honeycomb 
     		Kitaev model from numerically exact quantum Monte Carlo simulations \cite{Nasu2014}.
		A thermal cross-over indicated by a system-size invariant peak at temperature $T\approx 0.6~K$ 
		(where $K$ is the strength of the Kitaev coupling) reveals the temperature scale
		at which the original spin degrees of freedom fractionalize into Majorana fermions and a \Z\ gauge field. The latter
		undergoes an ordering transition indicated by the (diverging) lower-temperature peak at about 
		$T\approx 0.004~K$.
		Figure adapted from Ref.~\cite{Nasu2014}.
     		}
     \label{Fig:FiniteTemperatureSimulations}
\end{figure}

Three-dimensional Kitaev models distinguish themselves from their two-dimensional counterparts not only with regard to the variety of possible Majorana metals, but also with regard to the physics of their underlying \Z\ gauge theory. One striking difference between two and three spatial dimensions arises when considering the effect of thermal fluctuations on the order of the \Z\ gauge field. In two spatial dimensions such thermal fluctuations immediately melt the zero-temperature order of the \Z\ gauge field, while for three spatial dimensions it takes a critical strength of the thermal fluctuations to destroy the \Z\ order, i.e.\ there is a {\em finite-temperature} transition 
separating a low-temperature \Z\ ordered state from a high-temperature disordered state. The origin for the absence/occurrence of such a finite-temperature transition in two/three spatial dimensions can readily be understood when considering the nature of defects in the low-temperature \Z\ gauge order. In two spatial dimensions, e.g.\ for the honeycomb lattice, the elementary excitations are point-like vison excitations (associated with the plaquettes of the honeycomb lattice) that can freely move through the lattice (at no additional energy cost) and thereby 
thread a flux line through the system which destroys the long-range order of the \Z\ gauge field. For three spatial dimensions, the elementary vison excitations form closed flux {\em loops}, which leads to a competition between their excitation energy and configurational entropy. As a consequence, it takes a finite temperature to drive the system out of a regime of short flux loops into a regime of extended flux lines, which is connected to the paramagnetic high-temperature regime. 
In the context of the Kitaev model, this \Z\ gauge physics has been nicely demonstrated  \cite{Nasu2014} via numerically exact quantum Monte Carlo simulations (in the sign-problem free Majorana basis) of the Kitaev model on the hyper-honeycomb (10,3)b lattice. These finite-temperature simulations in fact reveal not only the \Z\ gauge transition at a temperature scale of about $T\approx 0.004~K$ (where $K$ is the strength of the Kitaev coupling), but also a thermal cross-over at around $T\approx 0.6~K$ (\ie on the bare scale of the Kitaev coupling), which indicates the actual spin fractionalization into Majorana fermions and a \Z\ gauge field, see the specific heat measurements illustrated in Fig.~\ref{Fig:FiniteTemperatureSimulations}.\newpage

The conceptual understanding of three-dimensional Kitaev models has been further expanded by analytical calculations of the dynamical structure factor for the hyper-honeycomb Kitaev model \cite{Smith2015} relevant to neutron scattering experiments and the Raman response for both the hyper-honeycomb and stripy-honeycomb Kitaev model \cite{Perreault2015}. The effects of disorder \cite{Sreejith2016} and interactions \cite{Hermanns2015} have been discussed for some of these lattices with the latter allowing for the possibility of a spin-Peierls instability. For the hyper-honeycomb lattice, extensions of the Kitaev model such as a three-dimensional Heisenberg-Kitaev model \cite{Kimchi2014,Lee2014,Kimchi2014b,Lee2014b} and the $J K \Gamma$-model of Eq.~\eqref{eq:HKG-model} have been considered \cite{Lee2015}.\newline
Finally, we note that beyond the analytically tractable 3D Kitaev models on tricoordinated lattices other 3D Kitaev generalizations for arbitrary lattice geometries \cite{Kimchi2014} have been considered. 


\subsection{$\beta$-Li$_2$IrO$_3$ and $\gamma$-Li$_2$IrO$_3$}

\begin{figure}[t]
    \centering
     \includegraphics[width=\hsize]{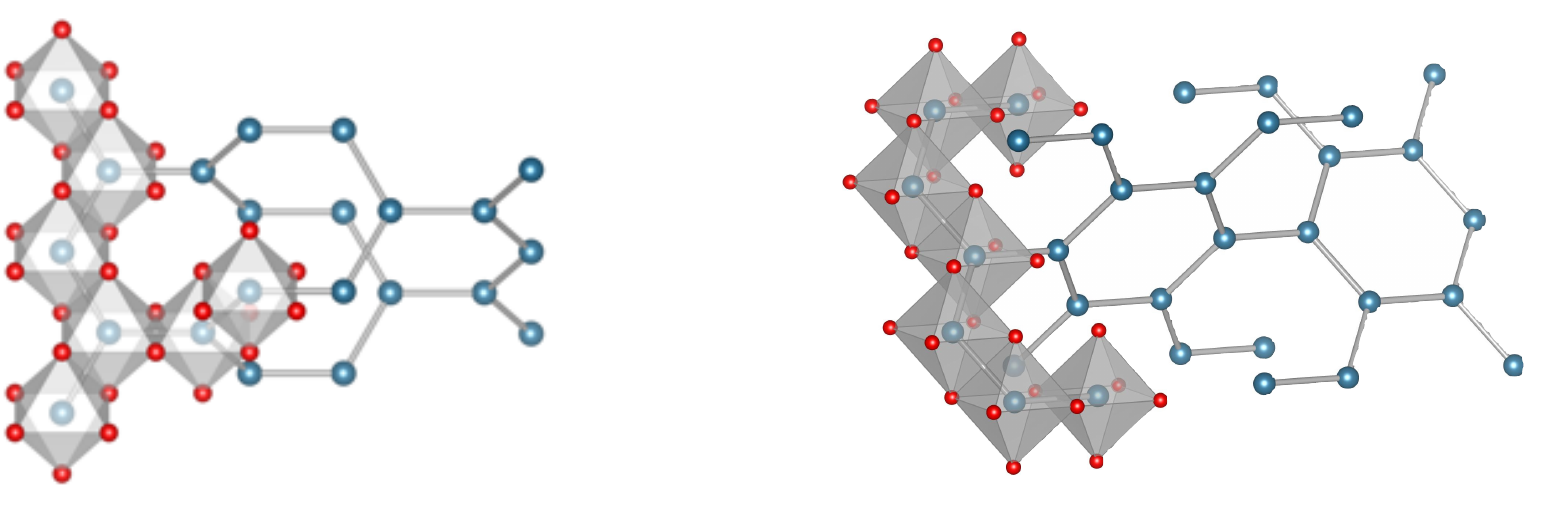}
     \caption{Crystal structure of the hyper-honeycomb Kitaev material \bLiIrO\ (left) and stripy-honeycomb Kitaev material \cLiIrO\ (right).}
     \label{Fig:3DIridates}
\end{figure}

The two  \LiIrO\ polymorphs \bLiIrO\ \cite{Takayama2015} and \cLiIrO\ \cite{Modic2014} are the first truly three-dimensional Kitaev materials -- realizing as illustrated in Fig.~\ref{Fig:3DIridates} a hyper-honeycomb and stripy-honeycomb lattice of egde-sharing IrO$_6$ octahedra, respectively. Independently synthesized almost at the same time, they are found to exhibit rather similar physics. 
Both systems are spin-orbit entangled Mott insulators with  effective moments of $1.6(1)~\mu_B$ close to the value of $1.73~\mu_B$ for an ideal $j=1/2$ moment. Susceptibility fits indicate  strong ferromagnetic interactions with estimates of the Curie-Weiss temperature found to be $\Theta_{\rm CW} \sim 40$~K ($\beta$) and $\Theta_{\rm CW} \sim 75$~K ($\gamma$), respectively. They both order at around $T_N \sim 38$~K into non-collinear magnetic order. Resonant magnetic x-ray diffraction  experiments \cite{Biffin2014a,Biffin2014b} on about $17-100~\mu$m wide single-crystals identify this non-collinear magnetic ordering with non-coplanar, counter-rotating long range spin spirals
with an incommensurate ordering wave vector $q = (0.57,0,0)$ along the orthorhombic $a$ axis in both materials. This unusual {\em counter-rotating} spiral order in these three-dimensional \LiIrO-polymorphs is thus very similar to the one (subsequently) observed in the original honeycomb material \aLiIrO.  

From a theoretical perspective, the two three-dimensional \LiIrO\ polymorphs have been investigated via ab initio calculations supporting the $j=1/2$ picture and dominant Kitaev-type bond-directional coupling \cite{Kim2015b,Katukuri2016}, which for the case of \cLiIrO, further argue
\cite{Li2017} for a reduced symmetry of the local Ir-O-Ir environment (possibly giving rise to rather complex magnetic interactions)  to explain the anisotropic behavior observed in optical conductivity measurements \cite{Hinton2015} for different polarizations. 
The origin of the magnetic ordering has been scrutinized based on a Heisenberg-Kitaev-Ising model \cite{Kimchi2014b,Kimchi2015} and $J K \Gamma$-model \cite{Lee2014,Lee2015,Lee2016} leading to a unifying theoretical framework \cite{Kimchi2015,Lee2016} for the spiral magnetism in all three \LiIrO\ polymorphs and their dynamics \cite{Kimchi2016}.

More recent experimental studies have argued for evidence of (Majorana) fermion quasiparticles in the Raman scattering \cite{Glamazda2016}, employing similar arguments as presented above in some detail for the Raman signatures of \RuCl, and the observation that a magnetic field with a small component along the magnetic easy-axis melts the magnetic long-range order, revealing a bistable, strongly correlated spin state \cite{Modic2016}. Future high-field experiments will be needed to assert whether this state is indeed a spin liquid. Along similar lines, it will be interesting to pursue high-pressure experiments, which have been argued to drive the three-dimensional Kitaev materials closer to the spin liquid regime \cite{Kim2016b}.


\subsection{Other materials}

We close by mentioning other materials scrutinized as three-dimensional Kitaev materials. This includes a recent theoretical suggestion  \cite{Masahiko2016} to synthesize metal-organic compounds such as honeycomb Ru-oxalate frameworks that might realize tricoordinated lattice structures in three spatial dimensions beyond the hyper-honeycomb (and its higher harmonics) such as the (10,3)a hyper-octagon lattice \cite{Hermanns2014}. Further, the Mott insulator La$_2$$B$IrO$_6$ ($B$=Mg,Zn) has been scrutinized \cite{Cook2015,Aczel2016} for its $j=1/2$ iridium moments being subject to a dominant Kitaev exchange on the face-centered cubic (fcc) lattice. 
Finally, also the hyperkagome material Na$_4$Ir$_3$O$_8$ \cite{Okamoto2007} has attracted some renewed interest \cite{Shindou2016,Kim2016} exploring the role of Kitaev-type interactions.

\newpage
\section{Outlook}

Before taking a look at the road ahead, it is very much worthwhile to note that in the few years since the original 2009 proposal for Kitaev physics in the $4d^5$ and $5d^5$ transition metals \cite{Jackeli2009} experimental progress has been made at an incredible pace. Not only have several  Kitaev materials with different two- and three-dimensional lattice geometries been synthesized and firmly established as $j=1/2$ spin-orbit entangled Mott insulators, but there has also been a streak of impressive experimental findings that most notably have provided direct evidence for bond-directional Kitaev-type interactions \cite{Kim2015} in the first Kitaev material \NaIrO\ and have firmly established the notion of a ``proximate spin liquid" in \RuCl. This includes the first experimental evidence of fermionic excitations in Raman scattering of a magnetic insulator \cite{Sandilands2015,Nasu2016} and a number of highly unusual signatures in inelastic neutron scattering (above the magnetic ordering transition) that can be well attributed to the proximity of spin liquid physics \cite{Banerjee2016,Banerjee2016b}. On the way, the unconventional magnetism of spin-orbit entangled $j=1/2$ moments in the \LiIrO\ polymorphs has been elucidated as a rare example of counter-rotating spin spirals \cite{Williams2016}.

As the field keeps moving with unflagging momentum, many novel materials are expected to take center stage in the coming years. Within the family of iridates, the hydrogen-intercalated honeycomb material $\alpha$-H$_{3/4}$Li$_{1/4}$IrO$_3$ \cite{Takagi2017} will undoubtedly receive much attention for its apparent spin liquid behavior. Beyond the iridates, other $5d^5$ oxides await experimental scrutiny such as the recently synthesized honeycomb rhodate Li$_2$RhO$_3$ \cite{Luo2013,Cao2013,Khuntia2015}. The next step might very well be to move beyond oxides altogether, which has already proved judicious with the exploration of the honeycomb chloride \RuCl. For instance, the potential for $j=1/2$ Mott insulators for iridium and rhenium {\em fluorides} has been highlighted in recent density functional and dynamical mean field theory studies \cite{Birol2015}. In further broadening the search for Kitaev materials, it might also be worthwhile to look beyond $4d$ and $5d$ transition metals and consider rare-earth magnets \cite{Li2016} whose $4f$ electrons are much more localized than the $5d$ or $4d$ electrons in iridates and ruthenates and at the same time experience a considerably stronger spin-orbit coupling -- thus  potentially providing another path to Kitaev materials in the future.

On a conceptual level, interest in Kitaev materials is spurred by their promise to realize what is a truly dichotomous state -- a spin-orbit entangled {\em Mott insulator}, in which the emergent degrees of freedom are Majorana fermions that form an (almost) conventional {\em metal}. These emergent Majorana metals exhibit distinct (topological) band structures \cite{Kitaev2006,OBrien2016} including the formation of Fermi surfaces, nodal lines, and Weyl nodes in three-dimensional settings along with the Dirac nodes of the originally proposed honeycomb system. 


\vskip 1cm

{\bf Acknowledgments}\newline
The author thanks his various collaborators on Kitaev physics over the years, first and foremost M. Hermanns along with B.~Bauer, M.~Becker, T.~Disselkamp, T.~Eschmann, K.~O'Brien, M.~Garst, M.~Gerlach, Z.-C.~Gu, H.-C. Jiang, V.~Lahtinen, A.~W.~W.~Ludwig, J.~K.~Pachos, X.-L. Qi, J.~Reuther, A.~Rosch, E.~Sela, and R.~Thomale, as well as P.~Gegenwart along with  experimental collaborators T.~Berlijn, W.~Ku, S.~Manni, and Y.~Singh. My group gratefully acknowledges support by the DFG within the CRC 1238 (projects C02, C03).


\bibliography{kitaev}

\end{document}